%% file: rev_1.tex
\documentclass[a4paper, 12pt]{article}
\newcommand{\la}{\langle}
\newcommand{\ra}{\rangle}
\renewcommand{\d}{\partial}
\newcommand{\beq}{\begin{eqnarray}}
\newcommand{\eeq}{\end{eqnarray}}
\newcommand{\sbeq}{\begin{subeqnarray}}
\newcommand{\seeq}{\end{subeqnarray}}
\newcommand{\eps}{\epsilon}

\newcommand{\btem}{\bibitem}
\newcommand{\YO}{Y. Oono}
\newcommand{\NGo}{N. Goldenfeld}

\newcommand{\bff}{\mbox{{\boldmath $f$}}}
\newcommand{\bfh}{\mbox{{\boldmath $h$}}}
\newcommand{\bfk}{\mbox{{\boldmath $k$}}}
\newcommand{\bfq}{\mbox{{\boldmath $q$}}}
\newcommand{\bfp}{\mbox{{\boldmath $p$}}}

\newcommand{\bfu}{\mbox{{\boldmath $u$}}}
\newcommand{\bfx}{\mbox{{\boldmath $x$}}}

\newcommand{\bfv}{\mbox{{\boldmath $v$}}}
\newcommand{\bfr}{\mbox{{\boldmath $r$}}}
\newcommand{\bfj}{\mbox{{\boldmath $j$}}}

\newcommand{\bfX}{\mbox{{\boldmath $X$}}}

\newcommand{\tf}{\tilde{f}}
\newcommand{\tbff}{\tilde{\bff}}
\newcommand{\tbfr}{\bar{\bfr}}

\newcommand{\bfR}{\mbox{{\boldmath $R$}}}
\newcommand{\bfJ}{\mbox{{\boldmath $J$}}}

\newcommand{\bfpsi}{\mbox{{\boldmath $\psi$}}}
\newcommand{\bfphi}{\mbox{{\boldmath $\varphi$}}}

\newcommand{\e}{{\rm e}}

\newcommand{\PRL}{Phys. Rev. Lett.}
\begin{document}
\begin{center}
\begin{Large}
{\bf Renormalization Group Method Applied to 
 Kinetic Equations: roles of initial values and time}
\end{Large}

\vspace{.5cm}
Y. Hatta$^1$ and T. Kunihiro$^2$
\end{center}
\begin{flushleft}
$^1$ Department 
of Physics, Kyoto University,  Kyoto 606-8502, Japan\\
$^2$ Yukawa Institute 
for Theoretical Physics, Kyoto University,
 Kyoto  606-8502, Japan
\end{flushleft}

\begin{abstract}
The so-called  renormalization group (RG) method is applied to 
derive kinetic and transport equations from the respective microscopic
 equations. The derived equations include Boltzmann equation in classical
 mechanics, Fokker-Planck equation, a rate equation in a quantum field
theoretical model. 
Utilizing the formulation of the RG method which elucidates the
 important role played by the choice of the initial conditions,
the general structure and the  underlying assumptions
 in  the derivation of kinetic equations in the RG method
is clarified.
It is shown that the present formulation naturally leads 
to the choice for the initial value of
 the microscopic distribution function at arbitrary time $t_0$ 
to be on the averaged distribution function to be determined.
The averaged distribution function may be thought as an integral constant
 of the solution of microscopic evolution equation;
the RG equation gives the slow dynamics of the would-be
 initial constant, which is actually  the kinetic equation
 governing the averaged distribution function.
It is further shown that the averaging as given  above
gives rise to a coarse-graining of
the time-derivative which is expressed with the initial time $t_0$,
 thereby leads to time-irreversible equations
even from a time-reversible equation.
It is shown that a further  reduction of  Boltzmann equation to fluid
dynamical equations and the adiabatic elimination of
 fast variables in Fokker-Planck equation are also performed in 
 a unified way in the present method.
\end{abstract}
\setcounter{equation}{0}
\renewcommand{\theequation}{\thesection.\arabic{equation}}

\section{Introduction}

Statistical physics, especially of non-equilibrium phenomena may be 
said to be  a collection of  theories on how to reduce the dynamics 
of many-body systems to ones with fewer variables\cite{kuramoto}.
BBGKY( Bogoliubov-Born-Green-Kirkwood-Yvon)
hierarchy\cite{reichl} which is equivalent to Liouville equation
hence time-reversible  can be reduced to the time-{\em ir}reversible
 Boltzmann equation\cite{boltzmann}
 given solely in terms of the single-particle
 distribution function for dilute gas systems\cite{bogoliubov}.
The derivation of Boltzmann equation by Bogoliubov\cite{bogoliubov}
 shows that  the dilute-gas dynamics as a {\em dynamical system}
with many-degrees of freedom 
has an {\em attractive manifold} \cite{holmes} spanned by 
 the one-particle distribution function,
 which is also an {\em invariant manifold} \cite{holmes}. 
Boltzmann equation in turn can be further
  reduced to the hydrodynamic equation (Navier-Stokes equation)
   by a perturbation theory like Chapman-Enskog 
  method\cite{chapman} or Bogoliubov's 
method\cite{bogoliubov,krylov}.
Langevin equation which may be time-irreversible can be
reduced to the time-irreversible Fokker-Planck equation
with a longer time scale than the scale in Langevin 
equation\cite{fp}.
Two basic ingredients are 
commonly seen in the reduction of dynamics, which are interrelated  
with  but relatively independent of each other:
(i) The reduced dynamics is characterized 
with a longer time scale than that appearing in the
original (microscopic) evolution equation.
(ii) The reduced dynamics is described by
 a time-irreversible equation even when the
 original microscopic equation is time-reversible.
The fundamental problem in the theory of deriving kinetic or transport
 equations is to clarify the mechanism of and to implement the above two
basic ingredients.

It seems that the basic notions to implement the two basic ingredients
(i) and (ii) are given by (1) the coarse-grained time-derivative 
and (2) the choice of the initial conditions in solving the
microscopic equations, respectively:\\ 
(1) In an attempt to characterize hydro-dynamical
processes microscopically,
 Mori pointed out that time derivatives appearing
in equations which define transport coefficients
 are ``the average'' of
time derivatives describing microscopic dynamics \cite{mori}. His
definition of the macroscopic derivative of an observable $F$ is 
\begin{eqnarray} 
\frac{\delta}{\delta t}\langle F\rangle(t)\equiv 
\frac{1}{\tau}\{\langle F\rangle (t+\tau)-
\langle F\rangle (t)\}=\frac{1}{\tau}\int ^{\tau}_0 ds 
\frac{d}{ds}\langle F\rangle (t+s),
\end{eqnarray}
where $\tau$ is some time scale between microscopic (mean free) 
time and
macroscopic (relaxation) time. An important point is that $\tau$ is
finite. 
The idea of the  coarse-graining 
of time in kinetic and transport equations were first given by
Kirkwood\cite{kirkwood}; see also \cite{ojima} for a rigorous
formulation.\\ 
(2) The importance of the choice of the 
initial condition in the derivation of kinetic equation is
noticed and  emphasized 
in the literature\cite{boltzmann,bogoliubov,kubo,lebowitz,kawasaki}.
For instance, 
 Kawasaki clarifies in an excellent monograph\cite{kawasaki} that
the initial value of the microscopic
distribution function before averaging must be given solely in
 terms of the averaged distribution 
to obtain a closed equation for the distribution function of
macroscopic slow variables, which is equivalent to the construction
of an invariant manifold mentioned above. He also clarifies that 
by this initial condition, the dominating class of states 
(``typical states'') are 
selected which
 leads to an increase of the entropy, while exceptional 
states from 
which the entropy would increase could be unstabilized to
 the dominating typical states by a mechanism producing chaotic 
behavior.
 Bogoliubov \cite{bogoliubov} also gave exactly the same 
scenario as described above in his asymptotic theory for the 
 reduction to Boltzmann equation;
 see chap. 9 of \cite{bogoliubov}.

In this paper, we apply the so-called renormalization group (RG)
\cite{RG} method  to derive and
 reduce kinetic equations to a slower dynamics
\cite{cgo,bricmont,kuni95,efk}:
We show that if the RG method is properly formulated 
 so that the essential role played by the choice of 
initial conditions is manifest as is done in \cite{kuni95,efk},
the general structure  in  the derivation of kinetic equations 
in the RG method is exactly in accordance with the basic principles 
(1) and (2) mentioned above.

The reduction-theoretical aspect of the RG method\cite{cgo} and the 
improved formulation given in \cite{kuni95} were
reformulated mathematically with the notion of
invariant manifolds familiar in the theory of  
 dynamical systems \cite{holmes} by Ei, Fujii and one of the present 
author\cite{efk}. In \cite{efk}, 
it was shown that the perturbative RG method can be used 
to construct invariant manifolds successively as the initial values 
of evolution equations using the Wilsonian RG \cite{RG,wilson,weg};
the would-be integral constants, which have one-to-one correspondence
 with the initial values, in the unperturbed
solution, constitute natural coordinates of the invariant manifold.
It was also shown that the RG equation determines the slow motion
of the would-be integral constants  on the invariant manifold of 
the dynamical system, hence a
reduction of evolution equation is achieved.

We shall show that the straightforward application of the RG method
as formulated in \cite{kuni95,efk} naturally leads 
to the choice for the initial value of
the microscopic distribution function at an arbitrary time $t_0$ 
to be on the averaged distribution, which is an implementation
of (2) in the RG method,
 thereby leads to time-irreversible equations
even from a time-reversible equation.
The averaged distribution function may be thought as an integral 
constant
 of the solution of microscopic evolution equation.
The RG equation gives the slow dynamics of the would-be
 initial constant, which is actually  the kinetic equation
 governing the averaged distribution function.
It will be  further shown that the averaging as given  above
automatically gives rise to a coarse-graining of
the time-derivative, which is expressed with the initial time $t_0$.
This shows that the initial time $t_0$ has a macroscopic nature
in contrast to the time $t$ appearing in the microscopic 
equations, which is an implementation of (1) in our method.

It should be noticed here that the RG equation has been already 
applied to kinetic equations both in
 classical and quantum regimes by others\cite{pashko,boyanovsky}:
In \cite{pashko}, it was ascertained that
 Boltzmann equation is a renormalization group equation  on the
 basis of the work by M.S. Green\cite{green} for the uniform system
 which shows that the perturbative solution for BBGKY hierarchy 
 exhibit a secular term,
and a sketch was given to derive
 Fokker-Planck equation from a simple Langevin equation
 noticing again an appearance of a secular term.
On the basis of  these two examples,
they claimed that all other   kinetic equations are
 also RG equations.
 Boyanovsky, de Vega and their collaborators\cite{boyanovsky}
have derived a Boltzmann equations for quantum field theories 
including gauge theories in the prescription of Illinois group
\cite{cgo}. They noticed that the existence 
of a mesoscopic scale which is microscopically large but
macroscopically small is essential for the applicability of 
the RG method to derive kinetic equations as well as the
validity of the description in terms of a kinetic equation 
itself; we remark here that the existence of such mesoscopic scales 
may be considered as the so-called intermediate asymptotics
\cite{barlenblatt}, which was actually the basic observation
 for the application
of the RG equation to asymptotic analysis of differential
 equations\cite{cgo}.
A model dealt in \cite{pashko,boyanovsky}
 will be retreated in our formulation, and implicit assumptions
in their treatment will be made explicit so that 
the roles of the initial conditions and the scale transformation
of the time-derivative will become clear for the RG method to
 lead to kinetic or transport equations.

In section 2, we shall deal with Langevin equation and derive the
Fokker-Planck equation as a typical problem of dynamical reduction
 leading to a kinetic equation in the RG method.
We shall  summarize the basic structure of the reduction given 
by the RG method.
One will see  that a similar definition to Eq.(1.1) of 
the macroscopic time-derivative  naturally emerges in the RG method.
In section 3,  Boltzmann equation is derived from Liouville
equation of the classical mechanics;
 we shall clarify the difference between 
the present method and the one by Bogoliubov.
In section 4, we apply
 the RG method to derive a Boltzmann equation(rate equation)
 in quantum field theory in a way where it is transparent
how  the initial
value are chosen and the coarse-grained time-derivative is
introduced in the RG method. This section is a
recapitulation of the work by Boyanovsky et al in our 
point of view.
In section 5, 
the RG method as formulated in \cite{efk} is applied to obtain
 the fluid dynamical limit of Boltzmann equation.
This is an example to  reduce a kinetic equation to a further slower
dynamics, which  appear quite often in statistical physics
 reflecting the hierarchy of the
space-time of the nature. Another typical problem in this category
 is to derive Smoluchowski equation from  Kramers
equation.
In section 6, we show that it is possible to develop a systematic 
theory based on the RG method 
for the adiabatic elimination of fast variables in
 Fokker-Planck equations.
Section 7 is devoted to a summary and concluding remarks.
In Appendix A, the calculational procedure sketched in \cite{pashko} 
for deriving Fokker-Planck equation in the RG method is
properly elaborated and worked out so that the essential role of 
the initial condition in the procedure is fully recognized;
it is shown that the initial value to obtain the 
``stochastic distribution function'' must be actually 
the averaged one for the RG equation to give Fokker-Planck 
equation in conformity with the observation by Kawasaki and 
others mentioned above.
%
\newpage
\setcounter{equation}{0}
\renewcommand{\theequation}{\thesection.\arabic{equation}}
\section{From Langevin to Fokker-Planck equation}  

In this section, we apply the RG equation to derive the 
Fokker-Planck(FP) equation\cite{fp}
 starting from the stochastic Liouville
equation \cite{kubo-s} corresponding to Langevin equation\cite{fp}.
This derivation is thought to be a typical one 
for the reduction of evolution equations appearing in
non-equilibrium physics\cite{kawasaki}.
It will be clarified that 
the initial values of the stochastic distribution function 
at arbitrary time $t_0$ are naturally chosen 
to be on the averaged distribution function
for the RG equation to derive the FP equation
 governing the averaged distribution function.
We shall also notice that the time derivative in the RG equation
which will be  converted to the derivative in the FP equation is
with respect to a macroscopic time, hence the coarse-graining of time
 is automatically built in in the present RG method.

\subsection{Application of the RG method 
to a generic equation Langevin equation}

We consider the following generic  Langevin equation with 
$R_i$ $(i=1, 2, ..., n)$ being stochastic variables;
\beq
\label{langevin:1}
\frac{d\bfu}{dt}=\bfh(\bfu)+\hat{g}(\bfu)\bfR,
\eeq
where $\bfu= \, ^t(u_1, u_2, ..., u_n)$,
$\bfh= \, ^t(h_1, h_2, ..., h_n)$, $\hat{g}$ a $n$ times $n$
matrix and $(\hat{g}(\bfu)\bfR)_i=\sum_jg_{ij}R_j$.
Notice that the noise enters multiplicatively.
Here we assume without loss of generality that 
the noise has  the vanishing average,
\beq
\la \bfR(t)\ra=0,
\eeq
where $\la {\cal O}(t)\ra$ denotes the average of ${\cal O}(t)$
 with respect to the noise $\bfR$.
Let $f(\bfu , t)$ be the distribution function with
$\bfR(t)$ given; the continuity equation reads
\beq
\label{continuity}
\frac{\partial f(\bfu ,t)}{\partial t}+\nabla_{_{\bfu}}\cdot
(\bfv f(\bfu ,t))=0,
\eeq
where $\nabla_{_{\bfu}}=\sum_i\partial/\partial u_i$ and 
 $\bfv=d\bfu/dt$ is the velocity of $\bfu$,
which is given in (\ref{langevin:1}).
Inserting (\ref{langevin:1}) into (\ref{continuity}), one has
 the Kubo's stochastic Liouville equation\cite{kubo-s},
\beq
\label{kubo}
\frac{\partial f}{\partial t}=-\nabla_{_{\bfu}}\cdot
              [(\bfh +\hat{g}\bfR)f].
\eeq
Whereas  it is rather easy task to  derive the FP equation in 
an exact way {\em if the noise is Gaussian}, it is formidably difficult 
if  the noise is {\em non-Gaussian}\cite{fp}.
 Although the present approach is based on 
the perturbation theory and of approximate nature, it will be
found that the first order calculation suffices to derive the
exact FP equation when the noise in Gaussian, and furthermore
that  the method is applicable even to non-Gaussian noises without
 difficulties.

The solution to (\ref{kubo}) with the initial
 condition given at $t=t_0$ is formally given by \cite{kubo-s}
\beq
\tilde{f}(\bfu ,t; t_0)=T{\rm exp}\bigl[\int_{t_0}^t ds L(s)\bigl]
\tilde{f}(\bfu ,t_0; t_0),
\eeq
where
\beq
L(s)=-\nabla_{_{\bfu}}\cdot(\bfh(\bfu)+\hat{g}\bfR(s)),
\eeq
 and $T$ denotes the time ordering operator. The initial distribution
$\tilde{f}(\bfu ,t_0; t_0)$ will be specified later and found to play a
significant role in the present method. 

Now we are interested in the averaged distribution function
$\tilde{P}(\bfu ,t; t_0)$ 
which is defined as an average of $f(\bfu ,t; t_0)$
with respect to the noise $\bfR$, i.e.,
\beq
\tilde{P}(\bfu ,t; t_0)=\la T{\rm exp}\bigl[\int_{t_0}^t ds L(s)\bigl]
f(\bfu , t_0)\ra.
\eeq
We take an interaction picture dividing the ``Hamiltonian''
$L$ as follows;
\beq
L&=&L_0+L_1, \\ 
 L_0&=& -\nabla_{_{\bfu}}\cdot\bfh, \quad
 L_1=-\nabla_{_{\bfu}}\hat{g}\bfR.
\eeq
We first define $U_0(t)$ as the time-evolution
operator governed by the unperturbed ``Hamiltonian'' 
$L_0$,
\beq
U_0(t,t_0)=T{\rm exp}\bigl[\int_{t_0}^t ds L_0(s)\bigl],
\eeq
which satisfies the evolution equation
\beq
\frac{\d}{\d t}U_0(t,t_0)=L_0U_0(t; t_0),
\eeq
with the initial condition
\beq
U_0(t_0, t_0)=1.
\eeq
Then to incorporate $L_1$, we define another microscopic distribution
 function $\rho_1(\bfu , t; t_0)$ by
\beq
f(\bfu ,t; t_0)=U_0(t,t_0)\rho_1(\bfu , t; t_0).
\eeq
We remark that the initial values of 
$f$ and $\rho_1$ at $t=t_0$ coincides with each other, which we take
 to be equal to the averaged distribution function $P(\bfu, t_0)$ 
at $t=t_0$ as depicted in Fig.1;
\beq
\label{init-value}
\tilde{f}(\bfu ,t=t_0; t_0)= \rho_1(\bfu ,t=t_0; t_0)=P(\bfu, t_0).
\eeq
One will recognize that this choice of the initial condition is 
inevitable for the RG equation to be identified with Fokker-Planck
 equation.
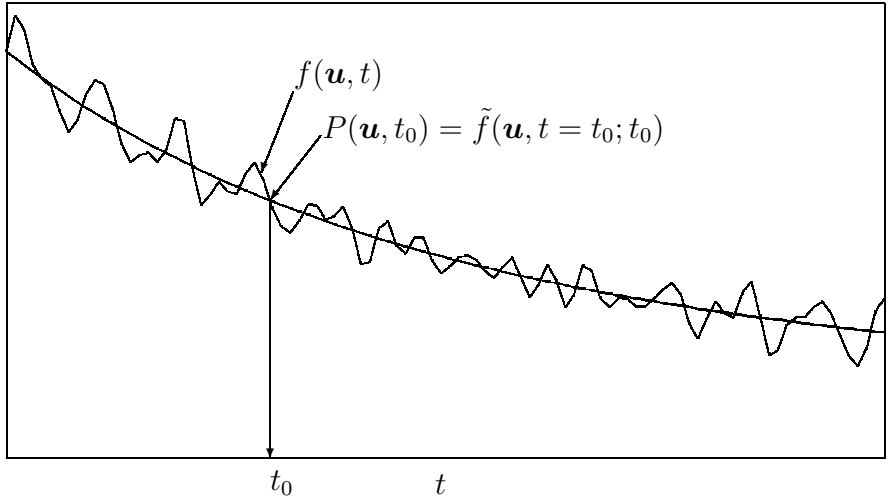
\begin{figure}
\begin{center}
\input{rg_fig}
\end{center}
\caption{The initial value of 
$\tilde{f}(\bfu ,t=t_0; t_0)= \rho_1(\bfu ,t=t_0; t_0)$ at $t=t_0$ 
is set on  the averaged distribution function $P(\bfu, t_0)$, which
 will be found to obey Fokker-Planck equation as the RG equation.
}
\end{figure}

One can easily verify that $\rho_1(\bfu , t; t_0)$ is
 formally solved to be
\beq
\rho_1(\bfu , t; t_0)=
T{\rm exp}\bigl[\int_{t_0}^t ds {\cal L}_1(s; t_0)\bigl]
\rho_1(\bfu , t_0; t_0),
\eeq
where
\beq
{\cal L}_1(t; t_0)=U_0^{-1}(t, t_0)L_1(t)U_0(t, t_0),
\eeq
is an ``interaction Hamiltonian'' in the ``interaction picture''.

Thus we obtains the compact form of $\tilde{P}(\bfu , t; t_0)$ as
follows,
\beq
\tilde{P}(\bfu ,t; t_0)&=& 
\la U_0(t, t_0)\rho_1(\bfu, t; t_0)\ra,\\ 
  \       &=& U_0(t, t_0)\la T{\rm exp}\bigl[\int_{t_0}^t ds 
              {\cal L}_1(s; t_0)\bigl]\ra P(\bfu, t_0),\\ 
          &\equiv &
              U_0(t, t_0)S(t; t_0)P(\bfu, t_0),
\eeq
where we have used the fact that 
$\rho_1(\bfu , t=t_0, t_0)=P(\bfu, t_0)$ and the abbreviation 
\[
S(t; t_0)\equiv
\la T{\rm exp}\bigl[\int _{t_0}^t ds{\cal L}_1(s; t_0)\bigl]\ra .
\]
The computation may be performed in a perturbative way:
\beq
\label{S-factor}
S(t; t_0)&=&
 1+ T\int _{t_0}^t ds \la{\cal L}_1(s)\ra+
\frac{1}{2}T\int _{t_0} ^t ds _1\int _{t_0} ^t ds _2
\la {\cal L}_1(s_1) {\cal L}_1(s_2) \ra  + \dots \nonumber \\ 
\ &=&  1+\frac{1}{2}T
\int_{t_0}^t ds_1\int_{t_0}^t ds_2\Gamma (s_1, s_2)
 + \dots \nonumber \\ 
\eeq
where we have put 
\beq
\Gamma (s_1,s_2)\equiv
\la {\cal L}_1(s_1){\cal L}_1(s_2)\ra.
\eeq
If the noise is stationary, which we shall assume from now,
$\Gamma(s_1, s_2)$ will be  a function of the difference 
$s_1-s_2$; furthermore,
owing to the time-reversible invariance of the
 microscopic law, $\Gamma (s_1, s_2)$ becomes a function of 
 the absolute value $\vert s_1-s_2\vert$, i.e.,
$\Gamma(s_1, s_2)=\Gamma (\vert s_1-s_2\vert)$.
Then one has for $t>t_0$,
\beq
S(t; t_0)=1+ (t-t_0)G(t-t_0)+\cdots,
\eeq
where we have put for $t>0$
\beq
G(t)=\int _{0}^t ds\Gamma(s).
\eeq
If we stop at the second order approximation,
we have 
\beq
\label{tildeP}
\tilde{P}(\bfu , t; t_0)=U(t; t_0)
                         \bigl[1+(t-t_0)G(t-t_0)]P(\bfu ,t_0).
\eeq
Notice the appearance of the secular term which indicates that 
the above formula is only valid for $t$ around $t_0$.

Now we apply the RG equation to (\ref{tildeP}) 
which  reads 
\[
 \partial \tilde{P}(\bfu , t;t_0)/\partial t_0\big\vert _{t_0=t}=0,
\]
which leads to
\[
\partial_{t_0}U_0(t, t_0)\vert_{t_0=t}P(\bfu, t)+
      \partial_tP(\bfu, t)-G(0)P(\bfu, t)=0,
\] 
where use has been made that $U_0(t_0, t_0)=1, \forall t_0$.
Noticing that 
$\partial_{t_0}U_0(t, t_0)\vert_{t_0=t}=-L_0=\nabla_{_{\bfu}}\cdot \bfh$,
 we arrive at Fokker-Planck equation,
\beq
\label{genericFP}
\partial_tP(\bfu , t)=-\nabla_{_{\bfu}}\cdot \bfh P(\bfu , t)
 +G(0)P(\bfu , t). 
\eeq
The concrete form of $G(0)$ depends on the 
character of the noise $\bfR(t)$.
This is one of the main results of this section.

To see that (\ref{genericFP}) is the desired equation,
let us evaluate $G(0)$ for a simple Gaussian noise given by
\beq
\label{gauss}
\la R_i(t)R_j(t')\ra=2\delta_{ij}D_i\delta(t-t').
\eeq
For this case, one has
\[
 \Gamma(s)=U_0^{-1}\partial_ig_{ij}
\partial_k g_{kl}2D_j\delta_{jl}\delta(s), 
\]
where $\partial_i=\partial/\partial u_i$.
Then $G(t)$ is evaluated as follows;
\beq
 G(t)&\equiv &\int_0^tds \Gamma(s)=
\frac{1}{2}U_0^{-1}\partial_ig_{ij}\partial_k g_{kl}2D_j\delta_{jl},
   \nonumber \\
  &=& G(0). 
\eeq
Here we have used the identity $\theta (0)=1/2$, in 
accordance with the Stratonovich scheme\cite{fp}. Notice that 
$G(t)$ in this case is independent of $t$.

Inserting $G(0)$ thus obtained into
 (\ref{genericFP}), one has the familiar form 
of Fokker-Planck equation for the 
multiplicative Gaussian noise,
\beq
\partial_tP(\bfu , t)=-\nabla_{_{\bfu}}\cdot \bfh P(\bfu , t)
            +D_j\partial_ig_{ij}\partial_kg_{kj}P(\bfu , t).
\eeq
This shows that the initial distribution $P(\bfu , t_0)$ 
satisfies Fokker-Planck equation and justifies the
 identification of the initial distribution  with the averaged 
one  made in Eq. (\ref{init-value}), as depicted in Fig. 1.

\subsection{Discussion}

Firstly, it is noteworthy  that we have been naturally led
to identify the initial values 
of the microscopic distribution function
$f(\bfu ,t_0, t_0)$ 
before averaging with the averaged value
 $P(\bfu ,t_0)$ at an arbitrary initial time $t=t_0$.
As mentioned in \S 1, the necessity to take such an initial condition
to achieve reduction of evolution equation was  advocated
 by Bogoliubov\cite{bogoliubov} and others\cite{lebowitz,kawasaki} 
including Boltzmann\cite{boltzmann}.
Secondly, this means that the nature of the initial time
$t_0$ in the RG method is completely different from that of the time
$t$ in the stochastic equation (microscopic
 equation); $t_0$ represents the coarse-grained time describing
 the variation of the averaged quantity, and the 
derivative $\partial_{t_0}$ in the RG equation 
is a macroscopic time-derivative.
Again as mentioned in \S 1, this coarse-graining
of time was also noted  by others \cite{kirkwood,mori,ojima}
 in different approaches.

This automatic averaging and the appearance of the
macroscopic time-derivative given in the RG method 
may be generically understood as a generalization of the scheme given
in \S 2 of \cite{efk}:
First discretize the variable $\bfu \rightarrow \bfu_i$
 and write as $P(\bfu , t)(\bfu _i , t)=X_i (t)$ and use a vector
notation $\bfX= (X_1, X_2, ...)$.
Thus the discretized  stochastic Liouville equation with the initial value
$\bfX (t_0)$
 at an arbitrary time $t_0$ may be solved perturbatively,
and the solution is denoted as 
$\tilde{\bfX}(t; t_0, \bfX(t_0))$, which
 satisfies the initial condition
\beq
\tilde{\bfX}(t_0; t_0, \bfX(t_0))=\bfX(t_0).
\eeq
We could solve the same equation with the initial condition
 given  at a shifted initial time $t=t_0+\Delta t$;
\beq
\tilde{\bfX}(t_0+\Delta t; t_0+\Delta t, \bfX(t_0+\Delta t))=
\bfX(t_0+\Delta t).
\eeq
 We suppose that the time difference
$\Delta t$ is  macroscopically small but 
microscopically so large that it may be taken as infinity.
For the time $t$ between $t_0$ and $t_0+\Delta t$, i.e.,
$t_0<t<t_0+\Delta t$, the perturbation should be valid.
 If $t-t_0$ and $\Delta t  \rightarrow \infty$  in the microscopic scale,
we may anticipate that the system is relaxed to the 
 averaged trajectory $\bfX(t)$ and have 
\[
\tilde{\bfX}(t; t_0+\Delta t, \bfX(t_0+\Delta t))\simeq
\tilde{\bfX}(t; t_0, \bfX(t_0)),
\]
which implies that the macroscopic time derivative
$\delta /{\delta t_0}$ vanishes,
\beq
0&=&\frac{\delta \tilde{\bfX}}{\delta t_0}\equiv
\frac{\tilde{\bfX}(t; t_0+\Delta t, \bfX(t_0+\Delta t))-
\tilde{\bfX}(t; t_0, \bfX(t_0))}{\Delta t_0}, \\
 &=& \frac{\partial\tilde{\bfX}}{\partial t_0}\vert _{_{t_0=t}}+
\frac{\partial \tilde{\bfX}}{\partial\bfX}\cdot
\frac{d{\bfX}}{dt_0}.
\eeq
Notice that in the macroscopic scale, the equality 
$t_0\simeq t\simeq t_0+\Delta t$ should be taken for granted.
This is the RG equation underlying the derivation of 
Fokker-Planck equation and also other transport equations
including kinetic equations as will be shown in the proceeding
sections.

In the following sections, we shall show that the general structure
elucidated in this section persists in other important examples
of the reduction to kinetic equations.
\newpage
\setcounter{equation}{0}
\renewcommand{\theequation}{\thesection.\arabic{equation}}
\section{From BBGKY hierarchy to  Boltzmann equation}

In this section, we apply  the RG method to derive  Boltzmann 
equation for a classical dilute gas starting from 
BBGKY hierarchy\cite{reichl}. 
As is well known,  Bogoliubov first derived  Boltzmann equation
 from BBGKY hierarchy in his classic
paper\cite{bogoliubov}. 
His derivation starts from an ansatz
that the many particle distribution function depends 
on time only through the one-particle distribution function and 
uses a special perturbative expansion method. 
His approach is actually an application and generalization 
of the asymptotic theory by Krylov and Bogoliubov (KB)
 successful to non-linear oscillators\cite{krylov}.
Here we do not make that ansatz and start from the {\em naive}
perturbation theory. We will see how the ansatz given by
 KB can be incorporated in the RG method. 
The importance of the initial condition again emerges, which 
is in accordance with the observation given in \cite{cohen}.
This implies that the appearance 
of a secular term\cite{pashko} does not constitutes the final story for the
derivation of Boltzmann equation.

\subsection{Derivation of  Boltzmann equation in the RG method}

Consider a system of $N$ identical classical particles enclosed in a
volume $V$. Following the notation of \cite{zubarev},
we denote the $i$-th particle's phase space coordinate by 
$x_i=({\bfr}_i,{\bfp}_i)$. The Hamiltonian of the system is
\begin{eqnarray}
H=\sum_{i=1}^N\frac{p_i^2}{2m} +
\frac{1}{2}\sum_{i\neq j} U(|{\bfr}_i-{\bfr}_j|).
\end{eqnarray}
We assume that the potential $U$ depends only on the relative distance
of two particles and that its range $d$ is much shorter than the mean
free path $l$.
The $N$-particle distribution function $f_N(x_1,\cdots,x_N,t)$ is
normalized as
\begin{eqnarray}
\int f_N(x_1,\cdots ,x_N,t)\frac{ \prod_{i=1}^N d{\bfr}_id{\bfp}_i}{N!}
=1.
\end{eqnarray}
We define the $s$-particle distribution function by 
\begin{eqnarray}
f_s(x_1,\cdots x_s,t)=\int f_N(x_1,\cdots ,x_N,t)\frac{dx_{s+1}\cdots 
dx_N}{(N-s)!}.
\end{eqnarray}
Then the normalization condition for $f_s$ becomes
 \begin{eqnarray}
\int f_s(x_1,\cdots,x_s,t)dx_1\cdots dx_s=\frac{N!}{(N-s!)}\simeq N^s,
\end{eqnarray}
from which we see that $f_s$ is of $s$-th order in the particle density
$n=\frac{N}{V}$. We assume that $n\ll 1$. 

The kinetic equation for $f_s$ is obtained by integrating Liouville
equation $\frac{d}{dt}f_N=0$ over $x_{s+1},\cdots,x_N$. Equations for
$f_1$ and $f_2$ read
\begin{eqnarray}
\frac{d}{dt} f_1(x_1,t)&=&(\frac{\partial}{\partial t} +iL_1^0)f_1(x_1,t)
=-\int dx_2L_{12}'f_2(x_1,x_2,t),  \label{1} \\
\frac{d}{dt}f_2(x_1,x_2,t)&=&
(\frac{\partial}{\partial t}  +iL_{12})f_2(x_1,x_2,t)
=-\int dx_3(iL_{13}'+iL_{23}')f_3(x_1,x_2,x_3,t) \label{2} ,
\end{eqnarray}
where
\begin{eqnarray}
L_{12}&=&L_1^0+L_2^0+L_{12}',\nonumber \\
L_i^0=-i\frac{{\bfp}_i}{m}\cdot \frac{\partial}{\partial{\bfr}_i},
& &\ \ \ \ \ \ 
L_{ij}'=i\frac{\partial U({\bfr}_i-{\bfr}_j) }{\partial{\bfr}_j}
\cdot (\frac{\partial}{\partial {\bfp}_j}-
\frac{\partial}{\partial {\bfp}_i}).
\end{eqnarray}
These are the first two equations of the 
BBGKY hierarchy which is a series of equations relating the 
evolution of $f_s$ to $f_{s+1}$. Our goal is to derive an equation 
(or equations)  which captures the essence of the
system's dynamics described by 
BBGKY hierarchy. In the language of
the the theory of dynamical systems\cite{holmes}, we wish to
 construct a low-dimensional invariant manifold
in the (practically) infinite-dimensional functional space 
spanned by $\{f_s\}$ and derive the reduced equations of motion 
 on it.

Whereas Liouville equation or,
equivalently, BBGKY hierarchy describes microscopic collisions
between particles in detail,  what interests us is the
macroscopic variation of the system caused by the accumulation of 
many collisions. More concretely, we wish to know the variation of 
the system over the space-time scale much longer than
 the collision radius and the
collision time and much shorter than the mean free path and the mean
free time. Such scale is called the {\it mesoscale}. The derivatives
appearing in (\ref{1}) and (\ref{2}) are, so to speak, microscopic
derivatives, while those
appearing in kinetic equations are macroscopic derivatives. We must
take into account their difference when deriving kinetic
equations.

Following \cite{efk}, 
suppose that we have found the solution to BBGKY hierarchy
$\{f_s(,t)\}$ up to an arbitrary time $t_0$. With the initial 
condition $\{f_s(,t_0)\}$ we try to solve (\ref{1}) and (\ref{2}) 
by the perturbative expansion in the density (virial expansion) to obtain a solution $\tilde{f}_s(t;t_0)$ around $t\sim t_0$.
Recalling that $f_s$ is of s-th order in the density, we expand as 
follows.
\begin{eqnarray}
\tilde{f}_1(x_1,t)&=&\tilde{f}_1^0(x_1,t)+\tilde{f}_1^1(x_1,t)+\tilde{f}_1^2(x_1,t)+\cdots,\\
\tilde{f}_2(x_1 ,x_2,t)&=&\tilde{f}_2^0(x_1,x_2,t)+\tilde{f}_2^1(x_1,x_2,t)+\cdots,\\
\tilde{f}_3(x_1,x_2,x_3,t)&=&\tilde{f}_3^0(x_1,x_2,x_3,t)+\cdots ,  
\end{eqnarray}
where $\tilde{f}_i^j(x_1,\cdots ,x_i,t)$ is of $(i+j)$-th order in the
density. Substituting the above expansion in (\ref{1}) and 
(\ref{2}), we get
\begin{eqnarray}
& &\frac{d}{dt} \tilde{f}_1^0(x_1,t)=0 \label{11}, \\
& &\frac{d}{dt}\tilde{f}_2^0(x_1,x_2,t)=0 \label{22}, \\
& &
(\frac{\partial}{\partial t} + 
\frac{{\bfp}_1}{m}\cdot \frac{\partial}{\partial{\bfr}_1} )\tilde{f}_1^1(x_1,t)
=\int dx_2\frac{\partial}{\partial {\bfr}_1} 
U(|{\bfr}_1-{\bfr}_2|)\cdot \frac{\partial}{\partial {\bfp}_1} 
\tilde{f}_2^0(x_1,x_2,t)\label{12},
\end{eqnarray}
where we have dropped terms which result in the surface integral.
We also expand the initial condition 
\begin{eqnarray}
f_1(x_1,t_0)&=&f_1^0(x_1,t_0)+f_1^1(x_1,t_0)+\cdots ,\nonumber \\
f_2(x_1,x_2,t_0)&=&f_2^0(x_1,x_2,t_0)+\cdots .
\end{eqnarray}
Equation (\ref{11}) and (\ref{22}) are easily integrated:
\begin{eqnarray}
\tilde{f}_1^0(x_1,t)&=&e^{-iL_1^0(t-t_0)}f_1^0(x_{1},t_0),\label{0}
 \nonumber \\
\tilde{f}_2^0(x_1,x_2,t)&=&e^{-iL_{12}(t-t_0)}f_2^0(x_{1},x_{2},t_0)=f_2^0(x_{10},x_{20},t_0),\label{abc}
\end{eqnarray}
where
\begin{eqnarray}
x_{i0}(x_1,x_2,t,t_0),\ \  \ i=1,2\label{xy}
\end{eqnarray}
are positions and momenta of the 
particles 1 and 2 at time $t_0$ under the influence of the 2-body
Hamiltonian 
\begin{eqnarray}
H^{(2)}\equiv \frac{p_1^2}{2m}+\frac{p_2^2}{2m}+U(|{\bfr_1}-
{\bfr}_2|).
\end{eqnarray}
The initial values $f_1^0(x_{1},t_0)$
 and $f_2^0(x_{1},x_{2},t_0)$ may be considered as
 the integration constants of the lowest-order equation.
In the RG method as formulated in \cite{kuni95,efk},
the integration constants will constitute 
 the coordinates of the zeroth invariant manifold\cite{holmes}.
The decisive step of the present approach is to choose the
initial condition as follows
\beq
f_2^0(x_{1},x_{2},t_0)= f_1^0(x_{1},t_0)f_1^0(x_{2},t_0),
\label{a2}
\eeq
irrespective of the distance between ${\bfr}_1$ and ${\bfr}_2$.
The underlying picture of this choice is 
 that the system is so  dilute that 
  the two particles at an arbitrary time $t_0$ are most probably 
located at distance much longer than the 
collision radius $d$, so that the correlation of the two 
particles is negligible and  $f_2$ can 
be set to the product of one-particle distribution functions. We 
remark that a probabilistic nature enters at this point\cite{landau}.

 The integration of (\ref{12})
 from $t_0$ to $t$ with $\frac{l}{v} \gg t-t_0$ 
($v$ is the average velocity), which implies that $t-t_0$
 is small in the macroscopic scale, gives
\begin{eqnarray}
\tilde{f}_1^1(x_1,t)&=&e^{-iL_1^0(t-t_0)}f_1^1(x_1,t_0)\nonumber \\ 
 & & +
\int_{t_0}^tdt'e^{-iL_1^0(t-t')}
\int dx_2\frac{\partial}{\partial {\bfr}_1} U(|{\bfr}_1-{\bfr}_2|)
\cdot \frac{\partial}{\partial
{\bfp}_1}f_1^0(x_{10}',t_0)f_1^0(x_{20}',t_0),
\label{ho}
\end{eqnarray}
where we have used (\ref{abc}) and (\ref{a2}),
and $x_{10}'$ and $x_{20}'$ are given by (\ref{xy}) with the
replacement $t\to t'$. 
We remark that the condition ($\frac{l}{v}\gg t-t_0$) 
is also 
required for the expansion in the density to be valid
\cite{bogoliubov}.
 In (\ref{ho}), 
only ${\bfr}_2$ for $|{\bfr}_1-{\bfr}_2|\le d$ contributes to the
integral. 
In this region, we can write 
\begin{eqnarray}
{\bfr}_{i0}'\sim {\bfr}_i-\frac{{\bfp}_{i0}}{m}(t'-t_0).
\end{eqnarray}
for a microscopically large period $t'-t_0\gg \frac{d}{v}$.
Here we have neglected vectors whose magnitudes are of order $d$.
 Then the perturbative solution {\it in the mesoscopic regime} 
$\frac{l}{v} \gg t-t_0\gg \frac{d}{v}$ is 
\begin{eqnarray}
\tilde{f}_1(x_1,t)&=&\tilde{f}_1^0(x_1,t)+\tilde{f}_1^1(x_1,t) \nonumber \\ 
 &=&e^{-iL_1^0(t-t_0)}f_1^0(x_1,t_0)
+\int_{t_0}^t dt'e^{-iL_1^0(t-t')}\int dx_2 \frac{\partial}{\partial {\bfr}_1} U(|{\bfr}_1-{\bfr}_2|) \\& & \cdot \frac{\partial}{\partial {\bfp}_{1}}f_1^0({\bfr}_1-\frac{{\bfp}_{10}}{m}(t'-t_0),{\bfp}_{10},t_0)f_1^0({\bfr}_2-\frac{{\bfp}_{20}}{m}(t'-t_0),{\bfp}_{20},t_0)\nonumber.\label{ff}
\end{eqnarray}
Note that ${\bfp}_{i0}={\bfp}_{i0}'$:
The magnitudes of
$\frac{l}{v}$ and $\frac{d}{v}$ are of course different for different
systems. For a dilute gas system, typical values are
 $10^{-8}\sim 10^{-9}$s and $10^{-12}\sim 10^{-13}$s, respectively.
The second term of the r.h.s. of (\ref{ff}) is the {\it secular
term}. Indeed, it can be shown that in the spatially homogeneous case it
is proportional to $t-t_0$ \cite{green}. Accordingly, we have chosen $f_1^1(,t_0)$ to be zero following the prescription given in \cite{efk}.
The RG equation reads
 \begin{eqnarray}
\frac{\partial}{\partial t_0} \tilde{f}_1(x_1,t)\biggl\vert_{t=t_0}=0\label{b} ,\\
\Rightarrow \ \ 
\frac{\partial}{\partial t}f_1^0(x_1,t)
&+&\frac{{\bfp}_1}{m}\cdot \frac{\partial}{\partial {\bf
r}_1}f_1^0(x_1,t)
\nonumber \\
&=&\int dx_2 \frac{\partial}{\partial {\bfr}_1} U(|{\bfr}_1-{\bf
r}_2|)\cdot 
\frac{\partial}{\partial {\bfp}_1}f_1^0({\bfr}_1,{\bfp}_{10},t)
f_1^0({\bfr}_1,{\bfp}_{20},t) \label{xr},
\end{eqnarray}
In (\ref{b}) we have imposed that $t=t_0$ although the expression
 (\ref{ff}) is valid for 
$t-t_0\gg\frac{d}{v}$ . 
This manipulation can be justified by the same logic given in the
 last part in \S 2 and will appear also in the case of field theory
 discussed in the following section:
The $t$-derivative is the microscopic derivative and 
the $t_0$-derivative
is the macroscopic one. Through the RG equation, we can
{\it automatically} go over to the mesoscopic physics from the
microscopic physics. 
Thus the mesoscopic nature of Boltzmann equation is transparent 
in our approach. We have also replaced the argument 
${\bfr}_2$ of $f_1^0$ with ${\bfr}_1$ in (\ref{xr}) 
because only $|{\bfr}_1-{\bfr}_2|\le d$ contributes to the
integration. 

(\ref{xr}) is the kinetic equation we have been seeking for. In the
language of the RG method, it is the renormalization group
equation describing the slow motion on the invariant manifold with the
coordinate $f_1^0(x_1,t)$ \cite{pashko}. To obtain the usual Boltzmann
equation which contains the gain minus loss term, we have to manipulate
the r.h.s. ignoring the spatial dependence. The result is 
\begin{eqnarray}
\frac{\partial}{\partial t}f_1^0(x_1,t)&+&\frac{{\bfp}_1}{m}\cdot
\frac{\partial}{\partial {\bfr}_1}f_1^0(x_1,t)\nonumber \\
&=&\int_0^{\infty} \rho d\rho \int_0^{2\pi}d\phi \int d{\bf
p}_2v_{12}\{f_1^0({\bfr}_1,{\bfp}_1',t)f_1^0({\bfr}_1,{\bf
p}_2',t)-f_1^0({\bfr}_1,{\bfp}_1,t)f_1^0({\bfr}_1,{\bf
p}_2,t)\} ,\nonumber \\
 & & 
\end{eqnarray}
where we have introduced the cylindrical coordinate pointing the
direction of the relative velocity ${\bfv}_{12}=({\bfp}_2-{\bf
p}_1)/m$. Comparing (\ref{11}) and (\ref{xr}), we see the change of the equation
by including the lowest-order contribution of the collision.

\subsection{Role of the initial condition}
Bogoliubov \cite{bogoliubov} 
was the first to point out that Boltzmann equation
represents the mesoscopic physics and derived it from Liouville
equation. 
In his derivation,
 he assumed that the system can be described only in terms of the one-particle
 distribution function. That is, starting from an arbitrary initial
 condition, the system will rapidly reach the state in which 
$f_s \ (s\ge 2)$ are functionals of  $f_1$.
\begin{eqnarray}
f_s(x_1,\cdots, x_s,t)=f_s(x_1,\cdots,x_s;f_1(\ ,t)) \ \ s\ge 2\label{a1}.
\end{eqnarray}
($f_s$ depends on time only through $f_1$.) 
In fact, this assumption is a basis of any kinetic theory.
 The special expansion method based on this assumption leads to a
special solution to BBGKY hierarchy. 
In the RG method, we started with a naive perturbative expansion
without any knowledge about kinetic theory. 
Physics enters when we choose a special initial condition (\ref{a2})
and with this choice we can construct an invariant manifold spanned by
the one-particle distribution function in the infinite-dimensional 
functional space, which was originally envisaged by 
Bogoliubov\cite{bogoliubov}; see also \cite{cohen}.
\newpage
\setcounter{equation}{0}
\renewcommand{\theequation}{\thesection.\arabic{equation}}

\section{Derivation of kinetic equation in quantum  field theory}

In this section we apply the RG method to 
derive Boltzmann equation 
for quantum field theory.
In fact, the application of the RG method
 to quantum field theory has been extensively done
by  Boyanovsky, de Vega and Wang \cite{boyanovsky} in the
prescription given by Illinois group\cite{cgo}.
In the present section,
 using the self-interacting scalar field theory,
 one of the model Lagrangians which Boyanovsky et al treated, 
we shall apply the RG method as formulated in \cite{kuni95,efk}
and elucidate how the choice of the initial values is 
essential to derive the time-irreversible equation and that the 
scale of the time in the RG equation is different
from the microscopic time in the original equation.
Although the formulation and the points to be emphasized 
 are shifted from what is given in \cite{boyanovsky}, we shall 
closely follow the calculation by Boyanovsky et al\cite{boyanovsky}.

Consider the following Hamiltonian:
\begin{eqnarray}
H&=&H'_0+H'_{int},\\
H'_0&=& \frac{1}{2}\int d^3x[\pi^2+(\nabla\phi)^2+m_0^2\phi^2]
                                    ,\nonumber \\
H'_{int}&=&\int d^3x\frac{\lambda}{4!} \phi^4.
\nonumber
\end{eqnarray}
At finite temperature, any particle feels the mean field produced
by the thermal excited particles which are present
even in the classical regime;  hence a change of the particle
picture from the bare one to what we call  the quasiparticle.
Fortunately, in the case of the scalar theory the medium (thermal) 
effect only affects the mass of the particle. 
So it is convenient to rewrite the Hamiltonian so that the
unperturbed one is written in terms of the quasiparticle 
operators as follows;
\begin{eqnarray}
H&=&H_0+H_{int},\\
H_{0}&=& \frac{1}{2}\int d^3x[\pi^2+(\nabla\phi)^2+
m_{eff}^2\phi^2]=
\int d^3k\omega_{\bfk}a^{\dagger}({\bfk})a({\bfk}) 
                                    ,\nonumber \\
H_{int}&=&\int d^3x[\frac{\lambda}{4!} \phi^4.
+\frac{1}{2}\delta m^2\phi^2]
\nonumber
\end{eqnarray}
\begin{eqnarray*}
\phi({\bfx},t)&=&\int \frac{d^3k}{(2\pi)^{3/2}}
\phi({\bfk},t)e^{i{\bfk}\cdot {\bfx}},\ \ \ \phi({\bfk},t)=
\frac{1}{\sqrt{2\omega_{{\bfk}}}} [a({\bfk},t)+
a^{\dagger}(-{\bfk},t)], \\
\pi ({\bfx},t)
&=&
\int \frac{d^3k}{(2\pi)^{3/2}} \pi({\bfk},t)
e^{i{\bfk}\cdot {\bfx}},\ \ \ \pi({\bfk},t)
=i\sqrt{\frac{\omega_{{\bfk}}}{2}} [a({\bfk},t)+
a^{\dagger} (-{\bfk},t)]\nonumber ,\\
\omega_{\bfk}&=&\sqrt{k^2+m_{\rm eff}^2}\label{y},
\ \ \ \ \ m_{\rm eff}^2+\delta m^2=m_0^2.
\end{eqnarray*}
Here
we have shifted the mass of the unperturbed Hamiltonian to the 
quasiparticle mass $m_{eff}$ which is to be 
determined self-consistently; in the equilibrium,
the result is nothing but the
effective mass to be given in the hard thermal loop resummation scheme 
\cite{braaten,boyanovsky};
\beq
m_{\rm eff}^2&=&m_0^2+\frac{\lambda}{2}\la \phi^2(x)\ra=
m_0^2+\frac{\lambda}{2}\int\frac{d^3k}{(2\pi)^3}
\frac{1+2n_{\bfk}}{2\omega_{\bfk}}, \quad
\omega_{\bfk}=\sqrt{k^2+m_{\rm eff}^2},
\eeq
where $n_{\bfk}$ is the Bose-Einstein distribution function;
accordingly, $\delta m^2=-\frac{\lambda}{2}\la \phi^2(x)\ra$.

For a non-equilibrium situation, 
the definition of the quasiparticles depends on time. 
We suppose that the distribution function 
$n_{\bfk}(t)\equiv n(t; \omega_{\bfk}(t))$ of the quasi
particles varies slowly. Here we have assumed that the system is spatially
 homogeneous,
so the evolution equation for the distribution function to be
obtained may be rather called the rate equation. 

To describe the slow variation,
 we anticipate that $n_{\bfk}(t)$
is a macroscopic distribution function varying slowly like
the distribution function $P(\bfu, t)$ 
governed by Fokker-Planck equation discussed in \S 2;
the corresponding counter part of the distribution 
function $\tilde{P}(\bfu, t; t_0)$ 
which includes rapid time-variations
 will be introduced shortly.

To solve the Heisenberg equation for $a({\bfk}, t)$,
 we impose the following  initial
condition at $t=t_0$ where $t_0$ is arbitrary;
\begin{eqnarray}
\langle a^{\dagger}({\bfk},t_0)a({\bfk'},t_0)\rangle 
\equiv n_{\bfk}(t_0)\delta ({\bfk}-{\bfk}'),
\end{eqnarray}
which implies that the quasi-particles has the mass dependent on 
the initial time $t_0$ as follows,
\beq
m_{\rm eff}^2(t_0)&=&m_0^2+\frac{\lambda}{2}\int\frac{d^3k}{(2\pi)^3}
\frac{1+2n_{\bfk}(t_0)}{2\omega_{\bfk}(t_0)}, \quad 
\omega_{\bfk}(t_0)=\sqrt{k^2+m_{\rm eff}^2(t_0)}.
\eeq

We define the microscopic distribution function around
 $t\sim t_0$ by
\begin{eqnarray}
\tilde{n}_{{\bfk}}(t; t_0, [n(t_0)])\delta({\bf 0})=
\langle a^{\dagger}({\bfk},t)a({\bfk},t)\rangle,
\end{eqnarray}
 with the initial condition
\beq
\tilde{n}_{{\bfk}}(t_0; t_0, [n(t_0)])=n_{\bfk}(t_0).
\eeq 
$\tilde{n}_{{\bfk}}(t; t_0, [n(t_0)])$
 is a counterpart of the microscopic distribution function
 $\tilde{P}(x, v, t; t_0)$ in contrast to $n_{\bfk}(t_0)$.
In the following, we shall suppress the third argument 
in $\tilde{n}_{{\bfk}}$ and write as 
$\tilde{n}_{{\bfk}}(t; t_0)$.
The equation of motion for the distribution function
$\tilde{n}_{{\bfk}}(t; t_0)$ 
is obtained by simply differentiating with respect to
 time and using the Heisenberg equation of motion.
The expectation value can be evaluated perturbatively 
by the closed time path formalism \cite{schwinger,keldysh,bellac}. 
To two loop order, the result is
\begin{eqnarray}
\frac{d}{dt}\tilde{n}_{{\bfk}}(t; t_0)=
\frac{\lambda^2}{3}\int^{\infty}_{-\infty}d\omega 
R(\omega,{\bfk};[n(t_0)])
\frac{\sin[(\omega -\omega_{{\bfk}})(t-t_0)]}
{\pi(\omega-\omega_{\bfk})}\label{rgeq},
\end{eqnarray}
\begin{eqnarray*}
R(\omega,{\bfk};[{n}(t_0)])
&\equiv&\frac{\pi}{2\omega_{\bfk}}
\int \prod_i \frac{d^3q}{(2\pi)^32\omega_{\bfq_i}}(2\pi)^3
\delta^3({\bfk}-{\bfq_1}-{\bfq_2}-{\bfq_3})\\
& \times&
\biggl\{ \delta(\omega+\omega_{\bfq_1}+\omega_{\bfq_2}+
\omega_{\bfq_3})N_1(t_0)+3\delta(\omega+\omega_{\bfq_1}+
\omega_{\bfq_2}-\omega_{\bfq_3})N_2(t_0)\\ 
&+&3\delta(\omega-\omega_{\bfq_1}-\omega_{\bfq_2}+
\omega_{\bfq_3})N_3(t_0)+\delta(\omega-\omega_{\bfq_1}-
\omega_{\bfq_2}-\omega_{\bfq_3})N_4(t_0)\biggr\},
\end{eqnarray*}
\begin{eqnarray*}
N_1(t_0)&=&(1+n_{{\bfk}}(t_0))(1+n_{{\bfq_1}}(t_0))
(1+n_{{\bfq_2}}(t_0))(1+n_{{\bfq_3}}(t_0))-n_{{\bfk}}(t_0)
n_{{\bfq_1}}(t_0)n_{{\bfq_2}}(t_0)n_{{\bfq_3}}(t_0),\\
N_2(t_0)&=&(1+n_{{\bfk}}(t_0))(1+n_{{\bfq_1}}(t_0))(1+n_{{\bfq_2}}(t_0))
n_{{\bfq_3}}(t_0)-n_{{\bfk}}(t_0)
n_{{\bfq_1}}(t_0)n_{{\bfq_2}}(t_0)(1+n_{{\bfq_3}}(t_0)),
\\ 
N_3(t_0)&=&
(1+n_{{\bfk}}(t_0))n_{{\bfq_1}}(t_0)
n_{{\bfq_2}}(t_0)(1+n_{{\bfq_3}}(t_0))-
n_{{\bfk}}(t_0)(1+n_{{\bfq_1}}(t_0))(1+n_{{\bfq_2}}(t_0))
n_{{\bfq_3}}(t_0),\\  
N_4(t_0)
&=&
(1+n_{{\bfk}}(t_0))n_{{\bfq_1}}(t_0)
n_{{\bfq_2}}(t_0)n_{{\bfq_3}}(t_0)-n_{{\bfk}}(t_0)(1+n_{{\bfq_1}}(t_0))
(1+n_{{\bfq_2}}(t_0))(1+n_{{\bfq_3}}(t_0)).
\end{eqnarray*}
Notice that the distribution functions in r.h.s are all the initial 
ones.
Integration over $t$ can be easily performed
\begin{eqnarray}
\tilde{n}_{\bfk}(t)=n_{\bfk}(t_0)+\frac{\lambda^2}{3}
 \int_{-\infty}^{\infty} 
d\omega R(\omega,{\bfk};[n(t_0)])
\frac{1-\cos[(\omega-\omega_{\bfk})(t-t_0)]}
{\pi(\omega-\omega_{\bfk})^2}.\label{nk}
\end{eqnarray}
The second term of r.h.s. can include
a secular term. Indeed, owing to 
the limiting behavior familiar in deriving the Fermi's golden rule
\begin{eqnarray}
\frac{1-\cos[(\omega-k)(t-t_0)]}{\pi(\omega-k)^2}\ \ 
\stackrel{t-t_0\to \infty}{\longrightarrow}(t-t_0)
\delta(\omega-k) \label{fermi},
\end{eqnarray}
 (\ref{nk}) becomes for $t-t_0 \to \infty$
\begin{eqnarray}
\tilde{n}_{\bfk}(t; t_0)
\sim n_{\bfk}(t_0)+\frac{\lambda^2}{3}(t-t_0)
R(\omega_{\bfk},{\bfk};[n(t_0)])+
\mbox{non-secular terms}.\label{se}
\end{eqnarray}
Notice that the r.h.s. is composed of the macroscopic 
distribution function.
The important point is that although we have taken the formal limit 
$t-t_0\to \infty$, 
(\ref{se}) is still 
a {\it local} expression valid {\it near} $t_0$. 
By ``near $t_0$'' we mean that $t-t_0$ is small in comparison with
 the macroscopic time scale which is 
in the same  order as the relaxation time.
Thus we are naturally led to suppose that
the time variable $t_0$ is a macroscopic one. 
The limit $t-t_0\to \infty$ merely means that $t-t_0$ is 
large in comparison with the microscopic time scale
which is  in the same order as 
the inverse of the (thermal) mass of the particle.
As was emphasized by Boyanovsky et al
 \cite{boyanovsky}, 
the separation of these scales is justified
 in the weak coupling regime, and the existence of such 
an intermediate scale is essential for the validity of 
the kinetic equation like Boltzmann equation.
A nice point in the RG method lies in the fact
 that we  naturally get to
 change our standpoint to observe the system from
 the {\it mesoscopic} point of view, as noticed above. 

Having made this change of the time scale,
the RG equation
\begin{eqnarray}
\frac{d}{dt_0}\tilde{n}_{\bfk}(t; t_0)
\biggl\vert_{t=t_0}&=&0\label{nn}
\end{eqnarray}
leads to the equation for the initial value $n_{\bfk}(t)$ as 
 the macroscopic variable with the
macroscopic time $t$ as follows
\begin{eqnarray}
\frac{d}{dt}n_{\bfk}(t)&=&
\frac{\lambda^2}{3}R(\omega_{\bfk},{\bfk};[n(t)])
=\frac{\lambda^2}{3}\frac{\pi}{2\omega_{\bfk}}
\int \prod_i \frac{d^3q}{(2\pi)^32\omega_{\bfq_i}}(2\pi)^3
\delta^3({\bfk}-{\bfq_1}-{\bfq_2}-{\bfq_3})
\nonumber \\ 
& \times&\biggl\{ \delta(\omega+\omega_{\bfq_1}+\omega_{\bfq_2}+
\omega_{\bfq_3})N_1(t)+3\delta(\omega+\omega_{\bfq_1}+
\omega_{\bfq_2}-\omega_{\bfq_3})N_2(t)
\nonumber \\ 
&+&
3\delta(\omega-\omega_{\bfq_1}-\omega_{\bfq_2}+
\omega_{\bfq_3})N_3(t)+\delta(\omega-\omega_{\bfq_1}-
\omega_{\bfq_2}-\omega_{\bfq_3})N_4(t)\biggr\}, \label{bolt}
\end{eqnarray}
which is nothing other than Boltzmann equation governing the
 slow variation of the distribution function. 
Thus the integration constant $n_{\bfk}(t)$ as given by the
 initial value has become a dynamical
 variable whose dynamics (\ref{bolt}) represents 
the reduced dynamics of the system, in accordance with the
general property of the RG method elucidated by Ei et al\cite{efk}.
We should remark that Eq.(\ref{bolt}) is supplemented with 
the time-dependent mean-field equation governing the 
time-evolution of $m_{\rm eff}(t)$,
 which is now given by 
\beq
m_{\rm eff}^2(t)&=&m_0^2+\frac{\lambda}{2}\int\frac{d^3k}{2\pi^3}
\frac{1+2n_{\bfk}(t)}{2\omega_{\bfk}(t)}, \quad
\omega_{\bfk}(t)=\sqrt{k^2+m_{\rm eff}^2(t)}.
\eeq
This condition leads to the r.h.s. of (4.8) which only contains the 
potential secular term and is equal to the time-dependent version of (4.3).
Thus the quasiparticle defined by the hard thermal loop resummation
 scheme is the relevant slow dynamical variable and we must choose the
 initial condition 
to be the distribution function of it to obtain the rate equation.
The importance of the choice of the initial condition  is
again recognized.
\newpage
\setcounter{equation}{0}
\renewcommand{\theequation}{\thesection.\arabic{equation}}
\section{Fluid dynamical limit of Boltzmann equation}  

In this section, 
we apply the RG method formulated above to obtain 
the fluid dynamical limit of Boltzmann equation\cite{boltzmann};
in other words, we shall derive Euler and Navier-Stokes
equations successively.

\subsection{Basics of Boltzmann equation}

Boltzmann equation\cite{boltzmann,resibois} is an evolution
 equation of the one-particle distribution function
$f(\bfr, \bfv, t)$ in the phase space, and reads
\beq
\frac{\d f}{\d t}+\bfv\cdot \frac{\d f}{\d \bfr}=I[f].
\eeq
Here the left-hand side
describes the change due to the canonical equation of 
motion while
 the right-hand side the change due to collisions;
\beq
I[f]&=&\int d\bfv_1\int d\bfv'\int d\bfv_1'w(\bfv\, \bfv_1
 \vert \bfv'\, \bfv_1')
      \nonumber \\ 
 \quad & & \times \biggl\{f(\bfr, \bfv', t)f(\bfr, \bfv_1', t)-
                   f(\bfr, \bfv, t)f(\bfr, \bfv_1, t)\biggl\} ,
\eeq
which is called the collision integral.
The transition probability
$w(\bfv\, \bfv_1 \vert \bfv'\, \bfv_1')$ has the following
 symmetry due to the time-reversal invariance of the
microscopic equation of motion;
\beq
\label{symm-1}
w(\bfv\, \bfv_1 \vert \bfv'\, \bfv_1')=
w(\bfv'\, \bfv_1' \vert \bfv\, \bfv_1).
\eeq
Furthermore, the invariance under the particle-interchange
 implies the following equality;
\beq
\label{symm-2}
w(\bfv\, \bfv_1 \vert \bfv'\, \bfv_1')=w(\bfv_1\, \bfv \vert \bfv_1'\, \bfv')
=w(\bfv_1'\, \bfv' \vert \bfv_1\, \bfv),
\eeq
where (\ref{symm-1}) has been used in the last equality.

Owing to the conservation  of the particle number,
the total momentum and the total kinetic energy during the
collisions, the following equalities hold,
\beq
\int d\bfv I[f]=0, \quad \int d\bfv \bfv I[f]=0, \quad
\int d\bfv v^2I[f]=0.
\eeq

We say that the function 
$\varphi (\bfv)$ is a {\em collision invariant} if 
it satisfies the following equation;
\beq \int d\bfv\varphi(\bfv)I[f]=0. 
\eeq
For a collision invariant $\varphi (\bfv)$,
 we can define the density $n_{\varphi}$ and the current
 $\bfj_{\varphi}$ as follows;
\beq 
n_{\varphi}=\int d\bfv
\varphi (\bfv)f(\bfr, \bfv, t), \quad \bfj_{\varphi}=\int d\bfv \bfv
\varphi (\bfv)f(\bfr, \bfv, t),
\eeq
which satisfy the continuity or balance equation;
\beq
\d_t n_{\varphi}+\nabla\cdot \bfj _{\varphi}=0.
\eeq

Thus we have formally the following 
fluid-dynamical equations as the balance equations for 
the conservation of the particle number,
 total momentum and kinetic energy;
\beq
\label{balance}
\d_t\rho +\nabla\cdot (\rho \bfu)&=&0, \\ 
\label{balance2}
\d_t(\rho u_i)+\d_j(\rho u_ju_i)+\d_j P_{ji}&=& 0, \\ 
\label{balance3}
\d_t(\rho\frac{u^2}{2}+e)+
\d_i\biggl[(\rho\frac{u^2}{2}+e)u_i+Q_i\biggl]
      +\d_i(P_{ij}u_j)&=&0,
\eeq
respectively.
Here, $\rho$, $\bfu$, $e$, $P_{ij}$, $Q_i$
 are the particle density, the stream velocity,
 the energy density, the pressure tensor and the 
heat flux,  respectively and defined as follows;
\beq
\rho (\bfr , t)&=&m\int d \bfv f(\bfr ,\bfv , t)=m\, n(\bfr, t),  \\
\rho (\bfr, t) \bfu (\bfr, t) &=&m\int d \bfv \bfv
 f(\bfr ,\bfv , t),
 \\ 
e(\bfr, t)&=&\int d \bfv \frac{m}{2}\vert \bfv - 
\bfu \vert ^2 f(\bfr, \bfv, t), \\
P_{ij}(\bfr, t)&=&\int d \bfv m(v_i -u_i)(v_j - u_j)f(\bfr, \bfv, t),
 \\
Q_i(\bfr , t)&=&\int d \bfv \frac{m}{2}\vert \bfv 
-\bfu\vert ^2 (v_i - u_i)
f(\bfr, \bfv , t).
\eeq
These equations are formal ones in the sense that 
the distribution function $f$ is not yet solved:
The solution obtained from Boltzmann equation 
will give  the explicit forms of the internal energy,
 the transport coefficients and so on.

The $H$ function is defined as follows:
\beq
H(\bfr, t)=\int d \bfv \, f(\bfr , \bfv , t)\big(\ln f(\bfr ,\bfv , t) -1\big).
\eeq
For equilibrium states, the $H$ function
 is equal to the entropy $S$ with the sign changed.
Defining the corresponding current by
\beq
\bfJ _H(\bfr , t)=\int d \bfv \, \bfv\,
 f(\bfr , \bfv , t)\big(\ln f(\bfr ,\bfv , t) -1\big),
\eeq
one has the balance equation;
\beq
\frac{\d H}{\d t}+\nabla\cdot \bfJ _H=\int d \bfv \, I[f]\ln f.
\eeq
This shows that when $\ln f$ is a collision invariant,
the $H$ function is conserved.
In fact, one can show that  this condition is satisfied when
$f(\bfr , \bfv , t)$ is a local equilibrium distribution
 function as given by (\ref{lmaxwell}) below.

\subsection{Derivation of fluid dynamical equations in the RG method}

To make it clear that the following discussion fits
 to  the general formulation given in \cite{efk},
 we discretize the argument $\bfv $ as 
$\bfv \rightarrow \bfv _i$\cite{kuramoto}:
 Discriminating the arguments $(\bfr , t)$ and $\bfv _i$ in
$f(\bfr, \bfv _i, t)$, we 
indicate $\bfv _i$ as a subscript $i$ for the 
distribution function;
\beq
f(\bfr, \bfv _i, t)=f_{i}(\bfr , t)\equiv (\bff (\bfr , t))_i.
\eeq
Then Boltzmann equation reads
\beq
\label{dbol-0}
\frac{\d f_{i}}{\d t} = \hat{I}[\bff]_i
- {\bfv}_i \cdot  \frac{\d f_{i}}{\d \bfr},
\eeq
where 
\beq
\hat{I}[\bff]_i = 
\sum_{j,k,l} w( {\bfv}_i{\bfv}_j\vert {\bfv}_k{\bfv}_l)
( f_kf_l - f_if_j )({\bfr}, t).
\eeq

Now let us consider a situation where
 the fluid motion is slow with long wave-lengths so that
\beq
\bfv _i\cdot \frac{\d f_i}{\d \bfr}=O(\eps), 
\eeq
where  $\eps $ is a small quantity,$ \vert \eps\vert <1.$

To take into account  the smallness of $\eps $ in the 
following calculations formally,
 let us introduce the scaled coordinate $\tbfr$ defined by
\beq
\tbfr =
\eps \bfr, \quad \frac{\d }{\d \bfr}=\eps \frac{\d }{\d \tbfr}. 
\eeq
Then (\ref{dbol-0}) now reads
\beq
\label{dbol-1}
\frac{\d f_{i}}{\d t} = \hat{I}[\bff]_i
- \eps {\bfv}_i \cdot  \frac{\d f_i}{\d \tbfr},
\eeq
which has a form to which the 
 perturbation theory given in \cite{efk} is naturally applicable.

In accordance with the general formulation given in \cite{efk},
 we first expand the solution as follows;
\[
f_i( {\tbfr},t ) = 
f_i^{(0)}( {\tbfr},t ) + \eps f_i^{(1)}( {\tbfr},t )+\cdots .
\]
Let $\tf _i(\tbfr , t; t_0))$ be a solution around
$t\sim t_0$ given by a perturbation theory with  
$f_i(\tbfr , t_0)$ being the initial value
 at $t=t_0$;
\beq
\tf _i(\tbfr , t=t_0; t_0)=f_i(\tbfr , t_0).
\eeq
We expand $\tf _i(\tbfr , t; t_0)$ as 
\beq
\tf_i( {\tbfr},t; t_0 ) = \tf_i^{(0)}( {\tbfr},t ; t_0 ) +
 \eps \tf_i^{(1)}( {\tbfr}, t, t_0 )+
\cdots, 
\eeq
and the respective initial condition
 is set up as follows,
\beq
\tf _i^{(l)}(\tbfr , t=t_0; t_0)=f_i^{(l)}(\tbfr , t_0), 
 \quad (l=0, 1, 2 ...).
\eeq

The $0$-th order equation reads
\beq
\frac{\d \tf _i^{(0)}}{\d t} = (I[{\tbff} ^{(0)}])_i.
\eeq
Now we are interested in 
 the slow motion which may be achieved 
 asymptotically as $t\rightarrow \infty$.
Therefore we take the following stationary 
 solution,
\beq
\frac{\d \tf _i^{(0)}}{\d t} = 0,
\eeq
which is  a fixed point of the 
 equation satisfying ,
\beq
\label{bol-fix}
(\hat{I}[{\tbff} ^{(0)}])_i=0,
\eeq
 for arbitrary $\tbfr$.
Notice that (\ref{bol-fix}) shows that 
the distribution function ${\tbff} ^{(0)}$ is a
 function of collision invariants.
Such a distribution function is a local equilibrium 
distribution function or Maxwellian;
\beq
\label{lmaxwell}
\tf_i^{(0)}(\tbfr , t; t_0 ) = 
n(\tbfr , t_0)
\biggl(
\frac{m}{2 \pi k_B T(\tbfr , t_0)}
\biggl)^{3/2}  	
\exp
\biggl[
- \ \frac{m \vert {\bfv}_i - {\bfu}(\tbfr ,t_0) \vert ^2}
{2 \pi k_B T(\tbfr ,t_0)}
\biggl].
\eeq
Here,
the local density $n$, local temperature $T$, local
 flux ${\bfu}$ are all dependent on
 the initial time $t_0$ and the space coordinate $\tbfr$
 but independent of time $t$.

The first-order equation reads
\beq
\biggl(
(\frac{\d}{\d t} - A){\tf}^{(1)}
\biggl)_i 
= - {\bfv}_i \cdot \frac{\d \tf_i^{(0)}}{\d {\tbfr}}.
\eeq
Here the linear operator $A$ is defined by
\beq
\biggl[\hat{I}'[\tbff ^{(0)}]{\tbff}^{(1)}\biggl]_i =
\sum_{j=1}^{\infty} \frac{\d I}{\d \tf _j}
\biggl\vert _{{\tbff}={\tbff}^0}\cdot \tf _j^{(1)} 
\equiv ( A{\tbff}^{(1)} )_i.
\eeq
The operator $A$ satisfies the following;
Let $ f_i =f_i^{(0)} \psi_i$. Then
\beq
(A \bff )_i=f_i^{(0)} 
\sum_{j,k,l} w( i j \vert k l ) f_j^{(0)} 
( \psi_k +  \psi_l - \psi_i- \psi_j )
\equiv f_i^{(0)}({\cal A}\bfpsi)_i.
\eeq
Here ${\cal A}$ is an operation acting on $\bfpsi$;
$\psi _i=(\bfpsi)_i$.
Defining the inner product between $\bfphi$ and $\bfpsi$ by
\beq
\la \bfphi, \bfpsi \ra =  \int d{\bfv} \varphi \psi, 
\eeq
one can show that $A$ is self-adjoint;
\beq
\la \varphi, A\psi\ra =\la A\varphi, \psi\ra.
\eeq
One can further show that 
the five invariants $m, \bfv , \frac{m}{2}{v}^2$
span the kernel of $A$\cite{resibois};
\beq
{\rm KerA} = \{ m,{\bfv},\frac{m}{2}{\bfv}^2 \}.
\eeq
The other eigenvalues are found to be 
negative because one can show
\beq
\la \varphi ,A \varphi \ra \quad  \le \quad   0.
\eeq

We write the projection operator 
 to the kernel as P and 
 define Q $= 1 -$ P.

Applying the general formulation in \cite{efk},
one can readily obtain the first-order solution,
\beq
\label{firstsol}
{\tbff^{(1)}} = 
- (t-t_0){\rm P}  {\bfv} \cdot 
\frac{\d \tbff ^{(0)}}{\d {\tbfr}}
+  A^{-1}{\rm Q}{\bfv} \cdot \frac{\d \tbff ^{(0)}}{\d {\tbfr}} .
\eeq
The perturbative solution up to the $\eps$ order is 
found to be 
\beq
{\tbff}(\tbfr , t, t_0)
= {\tbff}^{(0)}(\tbfr ,  t, t_0) + \eps [
- (t-t_0){\rm P}  {\bfv} \cdot 
\frac{\d \tbff ^{(0)}}{\d {\tbfr}}
+  A^{-1}{\rm Q}{\bfv} \cdot \frac{\d \tbff ^{(0)}}{\d {\tbfr}} ].
\eeq
Notice the appearance of a secular term.

If one stops the approximation and apply the
 RG equation,
${\d \tbff}/{\d t_0}\vert _{t_0 = t} = {\bf 0},$
 one has
\beq
\label{master:0}
 \frac{\d \tbff ^{(0)}}{\d t}  + 
\eps {\rm P} {\bfv} \cdot \frac{\d \tbff ^{(0)}}{\d \tbfr}={\bf 0}.
\eeq
This is a master equation from which equations governing
 the time evolution of $n(\bfr , t)$, $\bfu (\bfr , t)$
and $T(\bfr , t)$ in $\tbff ^{(0)}$,
 i.e., fluid dynamical equations:
In fact, taking an inner product between $m$, $m\bfv$ and 
$mv^2/2$ with this equation, one has fluid dynamical
 equations as given in (\ref{balance})-(\ref{balance3}).
 The only difference
 from (\ref{balance})-(\ref{balance3}) lies in the fact that
$f$ is now explicitly solved and 
 the energy density $e$,
 the pressure tensor $P_{ij}$ and the heat flux 
$Q_i$ are given as follows;
\beq
e(\bfr, t)&=&\int d \bfv 
\frac{m}{2}\vert \bfv - \bfu \vert ^2 f^{(0)}
(\bfr, \bfv, t)=\frac{3}{2}k_BT(\bfr , t), \\
P_{ij}(\bfr, t)&=&
\int d \bfv m(v_i -u_i)(v_j - u_j)f^{(0)}(\bfr, \bfv, t)=
  nk_BT(\bfr , t) \delta_{ij} \equiv P(\bfr , t) \delta_{ij}, \\
Q_i(\bfr , t)&=&
\int d \bfv \frac{m}{2}\vert \bfv -\bfu\vert ^2 (v_i - u_i)
f^{(0)}(\bfr, \bfv , t)=0.
\eeq
We have defined the pressure $P$ using 
 the equation of state for the ideal gas in the second
 line. There is no heat flux because the distribution 
function $f^{(0)}$ in the formulae is the one 
 for  the local equilibrium.
Inserting these formulae into (\ref{balance})-(\ref{balance3}),
we end up with a fluid dynamical equation without dissipation,
 i.e., the Euler equation in this approximation;
\beq
\frac{\d \rho }{\d t} + \nabla \cdot \rho \bfu &=&0, \\ 
\frac{\d (\rho u_i)}{\d t} + \frac{\d}{\d x_j} ( \rho u_i u_j ) + 
\frac{\d}{\d x_i} P &=& 0, \\
\frac{\d}{\d t}( \rho u^2 + e ) + \frac{\d}{\d x_i}
\biggl[ ( \rho \frac{u^2}{2} + e + P ) u_i \biggl] &=& 0. 
\eeq
We notice that these equations have been obtained from the
 RG equation (\ref{master:0}).
It should be emphasized however that the distribution function 
obtained in the present approximation takes the form
\beq
f(\bfr , \bfv , t)=f^{(0)}(\bfr , \bfv , t)+
A^{-1}{\rm Q}\bfv \cdot \frac{\d f^{(0)}(\bfr , \bfv , t)}{\d \bfr},
\eeq
which incorporates as a perturbation   
a distortion from the local equilibrium distribution and  gives
rise to dissipations.

One can proceed to the second order approximation straightforwardly
and  obtain fluid dynamical equation with dissipations as
 the RG equation.
The perturbation equation in the second order reads
\beq
\label{second}
\biggl(
(\frac{\d}{\d t} - A){\tf}^{(2)}
\biggl)_i 
= - {\bfv}_i \cdot \frac{\d \tf_i^{(1)}}{\d {\tbfr}}.
\eeq
Here, we must make an important notice:
We have actually used the linearized Boltzmann 
equation\cite{resibois}
 neglecting the second-order term of $\tbff ^{(1)}$ in 
the collision integral:
It is known that the neglected term produces  the so called
 Burnett terms which are absent in the usual Navier-Stokes
 equation \cite{landau}.

Inserting the first order solution (\ref{firstsol}) into {\ref{second}),
we have for the second order solution 
\beq
\bff ^{(2)}(\bar{\bfr}, t; t_0)&=&
-(t-t_0)\{A^{-1}{\rm Q}\bfv\cdot\nabla{\rm P}\bfv\cdot\nabla
+{\rm P}\bfv\cdot\nabla A^{-1}{\rm Q}\bfv\cdot\nabla \}
\bff^{(0)}(\bar{\bfr},t_0)\nonumber \\
 & & -\{(A^{-1}{\rm Q}\bfv\cdot\nabla)^2
 -\frac{1}{2}(t-t_0)^2({\rm P}\bfv\cdot\nabla)^2\}\bff ^{(0)}(\bar{\bfr},t_0).
\eeq
Adding up the all the solutions up to the second order and 
applying the RG equation
$d \tilde{f}(\bar{\bfr}, \bfv, t; t_0)/d t_0\vert_{t_0=t}=0$,
 we have
\beq
\frac{\d f^{(0)}}{\d t}&+&\eps\{ {\rm P}\bfv\cdot\nabla f^{(0)}
+A^{-1}{\rm Q}\bfv\cdot\nabla\frac{\d f^{(0)}}{\d t}
\}\nonumber \\
 &+&\eps^2 \{A^{-1}{\rm Q}\bfv\cdot\nabla{\rm P}\bfv\cdot\nabla+
{\rm P}\bfv\cdot \nabla A^{-1}{\bf Q}\bfv\cdot \nabla \}f^{(0)}=0.
\eeq
Notice that the time derivative of the distribution function 
in the first order terms in $\eps$ must be retained in this order,
  because
 the time derivative is at most of order $\eps$.
Applying the projection operators P and Q from the left,
we have
\beq
\frac{\d f^{(0)}}{\d t}+\eps {\rm P}\bfv\cdot\nabla f^{(0)}
+\eps^2 {\rm P}\bfv\cdot \nabla 
A^{-1}{\bf Q}\bfv\cdot \nabla f^{(0)}&=&0, \\
\eps A^{-1}{\rm Q}\bfv\cdot\nabla\{\frac{\d f^{(0)}}{\d t}+
\eps {\rm P}\bfv\cdot\nabla f^{(0)}\}&=& 0, 
\eeq
respectively.
Clearly, the second equation follows from the first one in this
order.
 The third term in the first equation represents dissipations.
Taking inner products between the first equation and
 the particle number, the velocity and
 the kinetic energy, one will obtain a fluid dynamical 
equation with dissipations included, i.e., Navier-Stokes 
equation\cite{resibois}.

In summary, we have shown that the fluid dynamical limit of the 
Boltzmann equation can be obtained  neatly in  the RG method
as formulated in \cite{efk}.
Such  problems to  reduce a kinetic equation to a further slower
dynamics appear quite often, reflecting the hierarchy of the
space-time of  nature. 
A typical problem 
\cite{fp,brinkman,wilemski,titulaer,kaneko,kampen2}
in this category
 is to derive Smoluchowski equation\cite{smoluchowski}
 from Kramers
equation\cite{kramers}.
As is expected, it is possible to develop a systematic 
theory for the adiabatic elimination of fast variables in
 Fokker-Planck equations, which is described in 
the next section.

\newpage
\setcounter{equation}{0}
\renewcommand{\theequation}{\thesection.\arabic{equation}}
\section{Adiabatic Elimination of Fast Variables in Fokker-Planck
Equation}

In this section, we shall show that the RG method as formulated in
\cite{efk} gives a systematic
way for the adiabatic elimination of fast variables appearing
 in Fokker-Planck equations
\cite{brinkman,fp,wilemski,titulaer,kaneko}.
A typical problem in this category is  to reduce
the so called Kramers equation\cite{kramers} to
 Smoluchowski equation\cite{smoluchowski}.
Although Brinkman\cite{brinkman} was the first for a serious attempt,
 reliable derivations was given rather lately
\cite{fp,wilemski,titulaer,kaneko}; see 
\cite{fp,kampen2} for a review.
Recently, it was shown that the RG method as formulated in 
\cite{cgo} can be used to eliminate a fast variable in the
Fokker-Planck equation\cite{matsuo}.
We shall show that the RG method
 gives the results which  have a clear correspondence  with those 
 given in other methods\cite{fp,kampen2};
thereby, we would say, our method gives a foundation for these methods.

\subsection{A generic FP equation with a fast variable}
Using  a generic example, we shall present 
the method for adiabatic elimination of fast variables
 based on the RG equation.

The example is  the following 2-dimensional 
Langevin equation with $\gamma$ being a large number,
\beq
\dot{x}&=&h_x(x, y)+g_x(x, y)\Gamma_x(t),\nonumber \\
\dot{y}&=&\gamma h_y(x, y)+f(x,y)+\sqrt{\gamma}g_y(x,y)\Gamma_y(t),
\eeq
where $\Gamma_i(t)$ ($i=x, y$) are independent
 Gaussian noise satisfying
\beq
\la \Gamma_i(t)\Gamma_j(t')\ra=2\delta_{ij}\delta(t-t').
\eeq
This example is treated by Gardiner\cite{fp}.
 We follow Gardiner for the notations.
We suppose that owing to the large friction $\gamma$,
the variable $y$ is a fast variable.
Our task is to eliminate the fast variable adiabatically
 and obtain the reduced dynamics for the slow variable $x$ only.
Before doing this task, we first eliminate the most rapid 
variables, i.e.,
 $\Gamma _i(t)$ ($i=x, y$)
 to obtain the dynamics written only in terms of
$x$ and $y$. This is tantamount to transforming the
Langevin equation to Fokker-Planck equation as was done in 
\S 2; it implies that what we are trying to is to further reduce a
kinetic equation to a slower dynamics. 
The corresponding Fokker-Planck equation for the 
probability $W(x, y, t)$ reads
\beq
\frac{\partial W}{\partial t}=[\hat{L}_1+\gamma \hat{L}_0]W,
\eeq
where
\beq
\hat{L}_1&=&-(\frac{\d}{\d x}D_x+\frac{\d}{\d y}f(x, y))+
\frac{\d^2}{\d x^2}D_{xx}, \\
\hat{L}_0&=&-(\frac{\d}{\d y}D_x+
\frac{\d^2}{\d y^2}D_{yy}),
\eeq
with
\beq
D_i=h_i(x,y)+g_i(x,y)\frac{d}{\d x_i}g_i(x, y),\quad
D_{ii}=g_i^2(x, y),   \quad (i=x, y).
\eeq

To apply the perturbation theory, we introduce a scaled time
$\tau $ by
\beq
\eps \tau=t,
\eeq
with $\eps=1/\gamma<1$ and write $\tau $ as $t$ for the moment.
Our equation now reads
\beq
(\partial_ t - \hat{L}_0)W=\eps\hat{L}_1W,
\eeq

 We denote the eigenvalues and the 
eigenfunctions of the unperturbed operator $\hat{L}_0(y, x)$
 as follows;
\beq
\hat{L}_0\varphi_n(y;x)=-\lambda_n(x)\varphi_n(y;x).
\eeq
Notice that $x$ in $\hat{L}_0(y; x)$ plays a role only as 
a parameter.
The corresponding adjoint equation reads
\beq
\hat{L}^{\dag}_0\varphi^{\dag}_n(y;x)=-\lambda_n(x)
\varphi^{\dag}_n(y;x).
\eeq
We assume that $\lambda_n(x)\geq 0$ and that the $0$-eigenvalue
$\lambda_0=0$ is non-degenerate:
The eigenfunction $\varphi_0(y;x)$ is proportional to
 the stationary
distribution function, so we can write as
$\varphi_0(y;x)=W_{st}(y;x)$.
It can be shown that the eigenfunctions $\varphi_n(y;x)$
and $\varphi^{\dag}_n(y;x)$ with
$n\geq 1$ are written in terms of $\varphi_0(y; x)$ as
\beq
\varphi_n(y; x)=\sqrt{\varphi_0(y; x)}\psi_n(y; x),\, 
 \quad \varphi^{\dag}_n(y; x)=\psi_n(y; x)/\sqrt{\varphi_0(y; x)},
\eeq
where $\psi_n(y; x)$ is an eigenfunction of the Hermite 
operator
\beq
\hat{H}_0=\frac{1}{\sqrt{\varphi_0(y; x)}}\hat{L}_0
\sqrt{\varphi_0(y; x)},
\eeq
with the eigenvalue $\lambda_n(x)$;
\beq
\hat{H}_0\psi_n(y; x)=-\lambda_n\psi_(y; x).
\eeq
Owing to the Hermiteness of $\hat{H}_0$,
the orthonormality and completeness hold;
\beq
\int\psi_n(y; x)\psi_m(y; x)dy&=&
\int\varphi^{\dag}_n(y; x)\varphi_m(y; x)dy=\delta_{n m},\\
\sum_{n=0}^{\infty}\psi_n(y; x)\psi_n(y'; x)&=&
\sum_{n=0}^{\infty}\varphi^{\dag}_n(y; x)\varphi_n(y'; x)
=\delta(y-y').
\eeq
For later use,
we define the projection operator $P$ 
to the kernel of $\hat{L}_0$ by
\beq
[P\varphi](y; x)=\varphi_0(y; x)\int \varphi^{\dag}_0(y'; x)
\varphi(y'; x)dy'.
\eeq
The projection operator to the orthogonal space to the kernel
is denoted as $Q$,
\beq
Q=1-P=\sum_{n=1}^{\infty}\varphi_n(y; x)\int dy'
\varphi^{\dag}_n(y'; x).
\eeq
We remark  that
\beq
\varphi^{\dag}_0(y; x)=1.
\eeq
In the following, we assume that 
\beq
\label{normal}
P\hat{L}_1P=0,
\eeq
which can be always satisfied by redefinition of $\hat{L}_0$ and 
$\hat{L}_1$. In terms of the notions in quantum field theory,
$\hat{L}_1$ is {\em normal-ordered} with respect to the vacuum
$\vert 0\ra\equiv \varphi_0$; $\la 0\vert \hat{L}_1\vert 0\ra=0$.

Let us suppose that a solution is given in the perturbation
 series;
$W(t, x, y)=W_0(t, x, y)+\eps W_1(t, x, y)+ \eps^2 W_2(t, x, y)$
$+\dots$.
Then following the general scheme
 of the RG method as given in Appendix A of \cite{efk},
we first construct the perturbed
solution 
\beq
\tilde{W}(t, x, y; t_0)=\tilde{W}_0(t, x, y;t_0)+
\eps \tilde{W}_1(t, x, y)+ \eps^2 \tilde{W}_2(t, x, y; t_0)
+\dots.
\eeq
 with the initial conditions given at arbitrary time
 $t=t_0$;
\beq
\tilde{W}_i(t_0, x, y; t_0)={W}_i(t_0, x, y).
\eeq
Notice that this implies that
\beq
W(t, x, y)=\tilde{W}(t, x, y; t).
\eeq
Now the equations for $\tilde{W}$ ($n=0, 1, 2, \dots$)  read
\beq
\label{eq0}
(\partial_ t- \hat{L}_0)\tilde{W}_0&=&0,\nonumber \\
\label{eqi}
(\partial_ t -\hat{L}_0)\tilde{W}_n&=&\hat{L}_1\tilde{W}_{n-1},
\quad (n=1, 2 , \dots).
\eeq

The solution to the lowest order equation around 
the initial time $t=t_0$ now reads
\beq
\tilde{W}_0(t,x,y; t_0)=\sum_{n=0}^{\infty}
C_n(x; t_0)\varphi_n(y; x)\e^{-\lambda_n t}.
\eeq
Notice that the coefficient function $C_n(x; t_0)$
 may depend on the initial time $t_0$ as well as $x$.
This solution implies that the initial value is taken as
\beq
W(t_0, x, y)=\tilde{W}_0(t_0,x,y; t_0)
  =\sum_{n=0}^{\infty}C_n(x; t_0)\varphi_n(y; x)\e^{-\lambda_n t_0}.
\eeq
If we are only concerned with 
 the asymptotic (long time) behavior as $t\rightarrow +\infty$, 
 however,
we may keep only  the stationary solution in the sum, namely,
\beq
\tilde{W}_0(t, x,y; t_0)=C_0(x; t_0)\varphi_0(y; x)=W_0(t_0, x, y).
\eeq
Notice the last equality, which gives the 
 functional form for the true solution in the $0$-th order,
holds because of the stationarity 
of the solution. We remark also that
\beq
\label{pvac}
P\tilde{W}_0(t, x,y; t_0)=\tilde{W}_0(t, x,y; t_0).
\eeq

Let us solve Eq.(\ref{eqi}) order by order.
The solution to Eq.(\ref{eqi}) is formally given by
\beq
\tilde{W}_n(t, x, y; t_0)&=&\frac 1{\d _t
-\hat{L}_0}\hat{L}_1\tilde{W}_{n-1}=
\frac 1{\d _t-\hat{L}_0}(P+Q)\hat{L}_1\tilde{W}_{n-1},\\
 &\equiv& X_n(t, x, y; t_0)+Y_n(t, x, y; t_0),  
\eeq
where
\beq
X_n&\equiv& P\tilde{W}_n(t, x, y; t_0)=
\frac 1{\d _t-\hat{L}_0}P\hat{L}_1\tilde{W}_{n-1},\nonumber \\
 Y_n&\equiv& Q\tilde{W}_n(t, x, y; t_0)=\frac 1{\d
 _t-\hat{L}_0}Q\hat{L}_1\tilde{W}_{n-1},
\eeq
with $n=1, 2, ...$
To obtain the special solutions to these
inhomogeneous equations,
one must impose an initial condition.
We impose the initial condition at $t=t_0$ so that 
 possible $P$-space components in the
 perturbed solutions is ``renormalized away'' to 
the unperturbed solution\cite{kuni95,efk}; 
namely, we shall construct special solutions which satisfy
\beq
\label{inicon}
P\tilde{W}_n(t_0, x, y; t_0)=X_n(t_0, x, y; t_0)=0.
\eeq
 We shall see that this condition is
 satisfied owing to the appearance of secular terms proportional
 to $(t-t_0)^n$ multiplied to the $P$ operator.

The special solutions to the perturbative equations (\ref{eqi}) with
the initial condition (\ref{inicon})
 may be obtained  in a mechanical way by the method given in Appendix
 A of\cite{efk}:
For any constant vector $U$
 the following formulae hold\cite{efk};
\beq
\label{formula1}
\frac 1{\d _t-\hat{L}_0}QU&=&\frac 1{-\hat{L}_0}QU,\\
\label{formula2}
\frac 1{\d _t-\hat{L}_0}(t-t_0)^nPU&=&\frac 1 {(n+1)}(t-t_0)^{n+1}PU,\\
\eeq
 for $n=0, 1, 2 ...$, especially for $n=0$, 
\beq
\label{formula3}
\frac 1{\d _t-\hat{L}_0}PU=(t-t_0)PU,
\eeq
and
\beq
\label{formula4}
\frac 1{\d _t-\hat{L}_0}(t-t_0)Q\hat{L}_1U&=&
(t-t_0)\frac 1{-\hat{L}_0}Q\hat{L}_1U-
 \frac 1{(-\hat{L}_0)^2}Q\hat{L}_1U.
\eeq
Here, notice that $[\hat{L}_0, Q]=0$, so 
$Q/\hat{L}_0={\hat L}_0^{-1}Q=Q{\hat L}_0^{-1}Q$.
We remark that these special solutions are all compatible 
with the initial condition (\ref{inicon}).

In the first order,  
one immediately sees that 
$X_1$ identically  vanishes because
\beq
\label{vanish}
P\hat{L}_1\tilde{W}_0=P\hat{L}_1P\tilde{W}_0=0,
\eeq
on account of (\ref{normal}) and (\ref{pvac}).
Thus we have for the first order solution
\beq
\label{1st}
\tilde{W}_1(t, x, y; t_0)=Y_1=\frac{1}{-\hat{L}_0}Q\hat{L}_1\tilde{W}_0.
\eeq
A couple of remarks are in order:
(i) One may replace the operator $Q$ with $P+Q=1$ in this expression
owing to  Eq.(\ref{vanish}).
Although such  simplification in the expressions
 may be done also for higher-order  solutions, we shall retain the
 $Q$ operator for definiteness, since the expression 
$(-\hat{L}_0)^{-1}\hat{L}_1$ itself is ill-defined because of the 
zero eigenvalue of $\hat{L}_0$.
(ii) The solution (\ref{1st}) satisfies the initial condition
(\ref{inicon}) as 
\beq
\tilde{W}_1(t_0, x, y, t_0)=W_1(t_0, x, y)=
\frac{1}{-\hat{L}_0}Q\hat{L}_1W_0(t_0, x, y).
\eeq

The second order terms are evaluated as follows:
\beq
X_2&=& \frac 1{\d
 _t-\hat{L}_0}P\hat{L}_1\frac{1}{-\hat{L}_0}Q\hat{L}_1\tilde{W}_0,\nonumber 
 \\
  &=& (t-t_0)P\hat{L}_1\frac{Q}{-\hat{L}_0}\hat{L}_1\tilde{W}_0,
\eeq
where (\ref{formula3}) has been used. Similarly, using (\ref{formula1})
we have
\beq
Y_2&=& \frac 1{\d _t-\hat{L}_0}Q\hat{L}_1
\frac{1}{-\hat{L}_0}Q\hat{L}_1\tilde{W}_0,\nonumber \\ 
 &=& (\frac{Q}{-\hat{L}_0}\hat{L}_1)^2\tilde{W}_0.
\eeq
Adding these terms, we have $\tilde{W}_2(t, x, y, t_0)$.
We remark that 
\beq
\tilde{W}_2(t_0, x, y, t_0)=W_2(t_0, x, y)=
 \frac{1}{\hat{L}_0}Q\hat{L}_1\frac{1}{\hat{L}_0}Q\hat{L}_1
W_0(t_0, x, y),
\eeq
which satisfies the initial condition (\ref{inicon}).

Let us proceed to the third order equation.
On account of (\ref{normal}), $P\hat{L}_1X_2=0$.
So,
\beq
X_3&=&\frac 1{\d _t-\hat{L}_0}P\hat{L}_1Y_2,\nonumber \\
 &=& (t-t_0)P\hat{L}_1(\frac{Q}{-\hat{L}_0}\hat{L}_1)^2\tilde{W}_0,
\eeq
where (\ref{formula3}) has been used in the last equality.
Similarly using (\ref{formula1}) and (\ref{formula4}), we have
\beq
Y_3&=&
\big[(t-t_0)\frac{Q}{-\hat{L}_0}\hat{L}_1P\hat{L}_1
\frac{Q}{-\hat{L}_0}\hat{L}_1
\nonumber \\
& &  -\frac{Q}{(-\hat{L}_0)^2}\hat{L}_1P\hat{L}_1\frac{Q}{-\hat{L}_0}\hat{L}_1
             + (\frac{Q}{-\hat{L}_0}\hat{L}_1)^3
                            \big]\tilde{W}_0.
\eeq
Adding the above terms, we have for the third order solution 
\beq
\tilde{W}_3(t, x, y; t_0)&=&
\big[(t-t_0)\{P\hat{L}_1(\frac{Q}{-\hat{L}_0}\hat{L}_1)^2
+\frac{Q}{-\hat{L}_0}\hat{L}_1P\hat{L}_1\frac{Q}{-\hat{L}_0}\hat{L}_1\}
\nonumber \\
& &
  -\frac{Q}{(-\hat{L}_0)^2}\hat{L}_1P\hat{L}_1\frac{Q}{-\hat{L}_0}\hat{L}_1
             + (\frac{Q}{-\hat{L}_0}\hat{L}_1)^3
                            \big]\tilde{W}_0,
\eeq
which satisfies the initial condition (\ref{inicon}).

Then utilizing the formulae (\ref{formula2}),
$X_4$ is evaluated to be
\beq
X_4&=&
\big[
\frac{1}{2}(t-t_0)^2P\hat{L}_1\frac{Q}{-\hat{L}_0}\hat{L}_1P\hat{L}_1
\frac{Q}{-\hat{L}_0}\hat{L}_1
+(t-t_0)\{-P\hat{L}_1\frac{Q}{(-\hat{L}_0)^2}\hat{L}_1P\hat{L}_1
\frac{Q}{-\hat{L}_0}\hat{L}_1\nonumber \\
 & & +P\hat{L}_1(\frac{Q}{-\hat{L}_0}\hat{L}_1)^3\}\big]\tilde{W}_0.
\eeq
Similarly with $Y_3$, we obtain 
\beq
Y_4&=&
\big[
(t-t_0)
\{\frac{Q}{-\hat{L}_0}\hat{L}_1P\hat{L}_1(\frac{Q}{-\hat{L}_0}\hat{L}_1)^2
+(\frac{Q}{-\hat{L}_0}\hat{L}_1)^2P\hat{L}_1\frac{Q}{-\hat{L}_0}\hat{L}_1    
          \}\nonumber \\
& & -\frac{Q}{(-\hat{L}_0)^2}\hat{L}_1P\hat{L}_1
(\frac{Q}{-\hat{L}_0}\hat{L}_1)^2
-\frac{Q}{-\hat{L}_0}(\frac{Q}{-\hat{L}_0}
\hat{L}_1)^2P\hat{L}_1\frac{Q}{-\hat{L}_0}\hat{L}_1
\nonumber \\ 
& &
-\frac{Q}{-\hat{L}_0}\hat{L}_1\frac{Q}{(-\hat{L}_0)^2}
\hat{L}_1P\hat{L}_1\frac{Q}{-\hat{L}_0}\hat{L}_1
+(\frac{Q}{-\hat{L}_0}\hat{L}_1)^4
                    \big]\tilde{W}_0.
\eeq
Adding the two terms, we have
\beq
\tilde{W}_4(t, x, y; t_0)&=&\big[
\frac{1}{2}(t-t_0)^2P\hat{L}_1\frac{Q}{-\hat{L}_0}\hat{L}_1P\hat{L}_1\frac{Q}{-\hat{L}_0}\hat{L}_1
 \nonumber \\
& &
+(t-t_0)\{-P\hat{L}_1\frac{Q}{(-\hat{L}_0)^2}\hat{L}_1P\hat{L}_1\frac{Q}{-\hat{L}_0}\hat{L}_1
+P\hat{L}_1(\frac{Q}{-\hat{L}_0}\hat{L}_1)^3
\nonumber \\
 & &
+\frac{Q}{-\hat{L}_0}\hat{L}_1P\hat{L}_1(\frac{Q}{-\hat{L}_0}\hat{L}_1)^2
+(\frac{Q}{-\hat{L}_0}\hat{L}_1)^2P\hat{L}_1\frac{Q}{-\hat{L}_0}\hat{L}_1    
          \}\nonumber \\
& & -\frac{Q}{(-\hat{L}_0)^2}\hat{L}_1P\hat{L}_1
(\frac{Q}{-\hat{L}_0}\hat{L}_1)^2
-\frac{Q}{-\hat{L}_0}(\frac{Q}{-\hat{L}_0}
\hat{L}_1)^2\hat{L}_1P\hat{L}_1\frac{Q}{-\hat{L}_0}\hat{L}_1
\nonumber \\ 
& &
-\frac{Q}{(-\hat{L}_0)^2}\hat{L}_1\frac{Q}{-\hat{L}_0}
\hat{L}_1P\hat{L}_1\frac{Q}{-\hat{L}_0}\hat{L}_1
+(\frac{Q}{-\hat{L}_0}\hat{L}_1)^4
                    \big]\tilde{W}_0.
\eeq
We remark that $P\tilde{W}_4(t_0, x, y; t_0)=0$.

If we stop at this order,
the perturbed solution is 
given by the sum 
\beq
\tilde{W}(t, x, y; t_0)\simeq
\sum_{i=0}^{4}\eps^i\tilde{W}_i(t, x, y; t_0).
\eeq
This solution becomes invalid as $t-t_0 \rightarrow \infty$
 due  to the secular terms.

Now the RG equation can now re-sum the secular terms:
The RG equation
\beq
\frac{d\tilde{W}}{dt_0}\biggl\vert_{t_0=t}=0
\eeq
gives for the reduced equations in the P- and Q-space
\beq
\label{rgW}
\d_tPW_0(t, x, y)=\eps^2P\hat{L}_1\big[
  \sum_{n=1}^3\eps^{n-1}(\frac{Q}{-\hat{L}_0}\hat{L}_1)^n
  -\eps^2\frac{Q}{-\hat{L}_0}\frac{Q}{-\hat{L}_0}\hat{L}_1P\hat{L}_1
      \frac{Q}{-\hat{L}_0}\hat{L}_1
\big]W_0(t, x, y),
\eeq
and
\beq
\label{qspace}
\{\eps\frac{Q}{-\hat{L}_0}\hat{L}_1+
\eps^2(\frac{Q}{-\hat{L}_0}\hat{L}_1)^2\}\d_t W_0 &=&
\{
\eps^3\frac{Q}{-\hat{L}_0}\hat{L}_1P\hat{L}_1\frac{Q}{-\hat{L}_0}\hat{L}_1
+\eps^4\frac{Q}{-\hat{L}_0}\hat{L}_1P\hat{L}_1
(\frac{Q}{-\hat{L}_0}\hat{L}_1)^2
 \nonumber  \\
 & &+\eps^4(\frac{Q}{-\hat{L}_0}\hat{L}_1)^2
 P\hat{L}_1\frac{Q}{-\hat{L}_0}\hat{L}_1\}W_0,
\eeq
respectively.
Here we have utilized the fact that 
$\tilde{W}_0(t, x, y; t_0=t)=W_0(t, x, y)$
and Eq.(\ref{pvac}).
It is readily verified that Eq.(\ref{qspace}) follows from 
 Eq.(\ref{rgW}).
Eq.(\ref{rgW}) is exactly the same as the one given 
by Gardiner\cite{fp} where the Laplace transformation was utilized.
The fourth-order term in (\ref{rgW}) may be rearranged 
as  a wave-function renormalization in this order 
of approximation;
\beq
\label{rgW2}
(1+\eps^2P\hat{L}_1\frac{Q}{(-\hat{L}_0)^2}\hat{L}_1)
\d_tW_0(t, x, y)=\eps^2P\hat{L}_1\big[
  \sum_{n=1}^3\eps^{n-1}(\frac{Q}{-\hat{L}_0}\hat{L}_1)^n
\big]W_0(t, x, y).
\eeq
In fact, multiplying Eq.(\ref{qspace}) by 
$-\eps P\hat{L}_1\hat{L}_0^{-1}$ and summing the result 
with Eq.(\ref{rgW}), one arrives at Eq.(\ref{rgW2}).

 Eq.(\ref{rgW}) or (\ref{rgW2}) is actually the equation for the amplitude
 $C_0(x, t)$:
Performing  the projection $P$ to the 
both sides, which is nothing but performing
the integration $\int dy$ of the both sides,
one has after recovering the original time $t\rightarrow t/\eps$
with $\eps=1/\gamma$,
\beq
(1+\gamma^{-2}\hat{\cal L}^{(N)}(x))\d _tC_0(x, t)=
\sum_{n=2}^4\gamma^{-(n-1)}\hat{\cal L}_n(x)
C_0(x, t),
\eeq
with
\beq
\hat{\cal L}^{(N)}(x)&=&\sum_{n=1}^{\infty}\hat{\cal L}_{0, n}
  \frac{1}{\lambda ^2_n(x)}\hat{\cal L}_{n, 0},\\
\hat{\cal L}_2(x)&=& \sum_{n=1}^{\infty}\hat{\cal L}_{0, n}
  \frac{1}{\lambda _n(x)}\hat{\cal L}_{n, 0},\\
\hat{\cal L}_3(x)&=& \sum_{n, m=1}^{\infty}\hat{\cal L}_{0, n}
  \frac{1}{\lambda _n(x)}  \hat{\cal L}_{n,m}
  \frac{1}{\lambda _m(x)}\hat{\cal L}_{m, 0}, \\
\hat{\cal L}_4(x)&=& \sum_{n, m, l=1}^{\infty}\hat{\cal L}_{0, n}
  \frac{1}{\lambda _n(x)}\hat{\cal L}_{n, m}
  \frac{1}{\lambda _m(x)}\hat{\cal L}_{m, l}
  \frac{1}{\lambda _l(x)}\hat{\cal L}_{l, 0}.
\eeq
Here $\hat{L}_{n,m}(x)$
are the $y$-averaged operators with respect to $x$,
defined by
\beq
\hat{\cal L}_{n, m}(x)= \int dy \varphi^{\dag}_n(y; x)\hat{L}_x(x, y)
   \varphi_m(y; x).
\eeq

\subsection{Derivation of Smoluchowski equation 
from a Kramers equation}

As an example, let us take the following simple case\cite{matsuo};
\beq
h_x(x, y)&=&y, \quad g_x(x, y)=0,\quad h_y(x, y)=-y,\nonumber \\
 f(x, y)&=& F(x),
\quad
 g_y(x, y)=\sqrt{T(x)}.
\eeq
This equation is a Langevin equation with a multiplicative
 noise.
In this case, we have
\beq
\hat{L}_0= \d_yy+T(x)\d^2_y, \quad 
\hat{L}_1=-y\d_x-f(x)\d_y.
\eeq
The eigenvalue problem for $\hat{L}_0$ is easily solved;
the eigenfunctions belonging to the eigenvalue 
$\lambda_n=n$, ($n=0, 1, 2 \dots$) may be written
\beq
\varphi_0(y; x)=\e^{-y^2/2T(x)}/\sqrt{2T(x)}, \quad 
\varphi_{n\geq 1}(y; x)=\frac{1}{2^nn!}H_n(y/\sqrt{2T(x)})\varphi_0(y; x),
\eeq
where $H_n(x)$ is the Hermite polynomial in the $n$-th order.
The conjugate eigenfunctions are found to be
\beq
\varphi^{\dag}_n(y; x)=H_n(y/\sqrt{2T(x)}),
 \quad \varphi^{\dag}_0(y; x)=1.
\eeq
Accordingly, the projection operator $P$ is given by a simple
integration,
\beq
P=\varphi_0(y; x)\int dy.
\eeq
To obtain the reduced dynamics, we only have to calculate the
$y$-averaged operators $\hat{\cal L}_{n, 0}(x)$ 
and $\hat{\cal L}_{0, n}(x)$.
First we notice that $P\hat{L}_1P=0$, hence
$\hat{\cal L}_1=\hat{\cal L}_{0, 0}(x)=0$. The non-vanishing term
 is evaluated to be
\beq
\hat{\cal L}_{0,n}(x)
&=&\int dy
 \hat{L}_1\varphi_n(y; x),\nonumber \\
 &=& -\int dy
 (y\d_x+f(x)\d_y)\varphi_n(y; x),\nonumber \\
 &=&-\delta_{1, n}\frac{\d}{\d x}\sqrt{\frac{T}{2}}.
\eeq
So we only have to calculate $\hat{\cal L}_{1, 0}$:
\beq
\hat{\cal L}_{1,0}(x)
&=&\int dy \varphi^{\dag}_1(y; x)
\hat{L}_1\varphi_0(y; x),\nonumber \\
&=&-\int dy \varphi^{\dag}_1(y; x)
(y\d_x+f(x)\d_y)\varphi_0(y; x).
\eeq
A straightforward calculation gives
\beq
(y\d_x+f(x)\d_y)\varphi_0(y; x)&=&\varphi_1(y; x)\sqrt{\frac{2}{T}}\{
T\frac{\d}{\d x} - f(x)+\frac{\d T}{\d x}\}\nonumber \\ 
   & & +\varphi_3(y; x)\sqrt{\frac{72}{T}}\frac{\d T}{\d x},
\eeq
hence
\beq
\hat{\cal L}_{1,0}(x)=
-\sqrt{\frac{2}{T}}\{T\frac{\d}{\d x} - f(x)+\frac{\d T}{\d x}\}.
\eeq
Using the above results, one obtains in the second order
approximation
\beq
\hat{\cal L}_2&=&\hat{L}_{0,1}\hat{L}_{1,0},\nonumber \\
              &=&\frac{\d}{\d x}\biggl(
T\frac{\d}{\d x} - f(x)+\frac{\d T}{\d x}
\biggl).
\eeq
Thus we end up with\cite{matsuo}
\beq
\label{rgc}
\frac{\d C_0(x,t)}{\d t}&=&\gamma ^{-1}\hat{\cal L}_2C_0(x; t), \nonumber \\
                        &=&\gamma ^{-1}\frac{\d}{\d x}\biggl(
T\frac{\d}{\d x} - f(x)+\frac{\d T}{\d x}
\biggl)C_0(x,t).
\eeq

One can proceed to higher orders.
 The third order correction is
found to vanish  owing to the  parity conservation
$\hat{\cal L}_{1,1}=\hat{\cal L}_{1,3}=0$.
So one must calculate the fourth order to have 
a correction to the second order result.
Although it is straightforward to calculate 
$\hat{\cal L}_4$ with $x$-dependent $T(x)$,
 we shall, in the following,  take a simple case where
$T$ is independent of $x$.
To obtain the fourth order correction,
one needs to evaluate the following operators:
\beq
\hat{\cal L}_{1, m}&=&-\delta_{m,2}\sqrt{\frac{T}{2}}\frac{\d}{\d x}
                    ,\\ 
\hat{\cal L}_{2, l}&=&\sqrt{\frac{T}{2}}\{\delta_{l, 3}\frac{\d}
{\d x} -\delta_{l, 1}4(\frac{\d}{\d x} - \frac{f(x)}{T})\},\\
\hat{\cal L}_{3,0}&=&0.   
\eeq
Thus one has
\beq
\hat{\cal L}_4=\frac{d^2}{\d x^2}(T\frac{\d}{\d x} -f)^2,
\eeq
which is exactly the same as given in other methods\cite{fp}.

\newpage
\section{Summary and Concluding Remarks}

In this paper, we have applied the so-called 
renormalization group(RG) method to derive and
 reduce kinetic equations; the equations treated include
 Boltzmann equation,
the fluid dynamical equation, Fokker-Planck equation for 
classical dynamics and also the rate equation in quantum
 field theory. In contrast to previous works\cite{pashko,boyanovsky},
 our main purpose was 
to elucidate the general structure of the reduction of
the dynamical equations in the hierarchy of the evolution equations.
We have  noticed that the significance of the choice of the 
initial value on the attractive manifold which is also an 
invariant manifold\cite{holmes} in deriving kinetic
equations is fully recognized and emphasized by 
Bogoliubov\cite{bogoliubov},
Lebowitz\cite{lebowitz}, Kubo\cite{kubo} and 
Kawasaki\cite{kawasaki}, for instance.
The notion of coarse-grained time derivative was also
 noticed by  Mori\cite{mori} and others\cite{kirkwood,ojima}.
Our point was that these basic ingredients naturally appear
 in the RG-theoretical derivation of kinetic equations when
 properly formulated so as to respect the role played by
the initial condition as formulated in \cite{kuni95,efk}.

We have also shown that the further reduction of
 kinetic equations can be performed in a unified manner in the 
 the RG method as formulated in \cite{efk}.
Such  problems include obtaining the fluid dynamical limit
 of Boltzmann equation and 
deriving  Smoluchowski equation from Kramers equation.

It is well known that the RG in quantum field theory(QFT)
and statistical physics\cite{RG,wilson,weg} 
works well as a powerful  tool for obtaining the infrared effective 
theories with fewer degrees of freedom than in 
the original Lagrangian relevant in the high-energy region. 
This is a kind of the reduction of the dynamics\cite{efk}. 
So one could  imagine that the RG may be 
applied to derive and reduce kinetic equations
 in a unified manner. Now we know that it is the case.
Conversely, does the coarse graining of time as shown in the 
present work appear  and play a role 
in the RG equations in QFT and statistical
 mechanics?  In this respect, one may notice the resemblance of
the exact RG equation like 
Wilsonian or Wegner-Houghton RG equation with Fokker-Planck
equation; the quantum filed theory with high-energy degrees of 
freedom remained corresponds to Langevin equation and
the exact RG equation to Fokker-Planck.
Such an analogy is not fully pursued yet\cite{jz}. 

\vspace{.5cm}
\begin{center}
\begin{large}
Acknowledgments
\end{large}
\end{center}
Part of the present work  was presented by T.K. at RIKEN-BNL workshop 
July 17-30, 2000, BNL.
T.K. is grateful to the organizers
 for inviting him to the workshop.
We thank T. Koide and Y. Nemoto for their discussions.
We acknowledge the members of the nuclear theory group in Kyoto university
for their interest in  this work.
We are grateful to the referee for useful comments to make the
exposition of the  paper more transparent.
T.K. is financially  supported by the Grants-in-Aid of
the Japanese Ministry of Education, Science and Culture
(No. 12640263 and 12640296).
\newpage
\setcounter{equation}{0}
\renewcommand{\theequation}{A.\arabic{equation}}

\begin{Large}
{\bf Appendix A \quad  Another calculational procedure for reduction
 of Langevin  to Fokker-Planck equation}
\end{Large}
\vspace{1cm}

In this Appendix, we shall rederive Fokker-Planck 
equation for (\ref{langevin:1}) starting
 from the stochastic Liouville equation in a more elementary 
way sketched in \cite{pashko} for far simpler equation;
 we shall work out a detailed
derivation to show how the identification of the initial
condition is important for obtaining the averaged equation,
thereby  exhibit clearly the similarity of the discussion
 with the other problems considered in the text.

We first make a change of independent variables\cite{pashko}
 for (\ref{kubo})
 $(t, \bfu ) \rightarrow (\tau , \bfx)$ by
\beq
\tau = t, \quad 
 \bfx =\eps\{ \bfu  -\int ^tds \bfh(\bfu(s))\},
\eeq
where $\eps $ is supposed to be small. 
Then Eq. (\ref{kubo}) is converted to
\beq
\label{modified-kubo}
\partial f/\partial \tau = -\eps
[\nabla(\hat{g}\bfR f)+\nabla\cdot \bfh f],
\eeq
where $\nabla=\sum _i\partial/\partial x_1.$ 
We now try to solve Eq.(\ref{modified-kubo}) around $t\sim \forall
t_0$ by the perturbation theory.
We suppose that the initial distribution $\bar{f}(\bfu , t_0)$
 is given at $t=t_0$. The corresponding solution is written as 
$\tilde{f}(\bfu , t; t_0)$ and is expanded as
\[
\tilde{f}=\tilde{f}_0+\eps\tilde{f}_1+\eps^2\tilde{f}_2+\cdots .
\]
The initial distribution is also expanded;
\[
\bar{f}=\bar{f}_0+\eps\bar{f}_1+\eps^2\bar{f}_2+\cdots .
\]

The equations in the first few orders read
\beq
\partial \tilde{f}_0/\partial \tau &=&0, \\
\partial \tilde{f}_1/\partial \tau& = &
-[\nabla(\hat{g}\bfR \tilde{f}_0)+\nabla\cdot \bfh \tilde{f}_0], \\ 
\partial \tilde{f}_2/\partial \tau &=& 
-[\nabla(\hat{g}\bfR \tilde{f}_1)+\nabla\cdot \bfh \tilde{f}_1].
\eeq
The first order solution is a stationary one;
\beq
\tilde{f}_0(\bfx , t; t_0)=\bar{f}_0(\bfx, t_0),
\eeq
where the initial distribution $\bar{f}_0(\bfx, t_0)$ is not yet 
specified.
Owing to the absence of explicit time-dependence of $\tilde{f}_0$,
the first and the second order equations can be readily solved;
\beq
\tilde{f}_1(\bfx , t; t_0)&=& -\int_{t_0}^tds\nabla\cdot(\hat{g}\bfR
\bar{f}_0) - (t-t_0)(\nabla\cdot \bfh)\bar{f}_0, \\
\tilde{f}_2(\bfx , t; t_0)&=&
 \int_{t_0}^tds_1\int_{t_0}^{s_1}ds_2 L_1(s_1)L_1(s_2)\bar{f}_0
 +\frac{1}{2} \nabla\cdot\bfh\nabla\cdot \bfh
 \bar{f}_0(t-t_0)^2\nonumber
    \\
  & &   + {\rm terms\, linear\, in}\, \bfR .
\eeq
Here $L_1(s)$ is defined in Eq.(2.9) in the text.
The averaged distribution function $\tilde{P}(\bfx ,t; t_0)$ is now
 given by 
\beq
\tilde{P}(\bfu , t; t_0)&=&\la \tilde{f}(\bfu , t; t_0)\ra,\nonumber \\ 
   &=& \bar{f}_0(\bfu, t_0)- \eps(t-t_0)(\nabla\cdot\bfh 
\bar{f}_0(\bfu , t_0)\nonumber \\
    & &  +
\eps^2 \int_{t_0}^tds_1\int_{t_0}^{s_1}ds_2\la L_1(s_1)L_1(s_2)\ra 
\bar{f}_0(\bfu , t_0)\nonumber \\ 
    & & 
 +\frac{1}{2}(t-t_0)^2 \nabla\cdot\bfh\nabla\cdot \bfh \bar{f}_0(\bfu , t_0).
\eeq
For steady noises, the correlation 
$\int_{t_0}^tds_1\int_{t_0}^{s_1}ds_2\la L_1(s_1)L_1(s_2)\ra $
 can be further reduced as was done in \S 2;
\beq
\int_{t_0}^tds_1\int_{t_0}^{s_1}ds_2\la L_1(s_1)L_1(s_2)\ra 
=(t-t_0)G(t-t_0).
\eeq

The RG equation $\partial \tilde{P}/\partial t_0\vert _{t_0=t}=0$
 gives 
\beq
{\d \bar{f}_0(\bfu ,t)}/\partial t+
\eps \nabla\cdot \bfh \bar{f}_0-\eps^2
G(0)\bar{f}_0(\bfu , t_0))
=0 .
\eeq
Recovering the original variables, we finally have
\beq
{\d\bar{f}_0}/\partial t+
 \nabla(\cdot \bfh \bar{f}_0)-
G(0)\bar{f}_0
=0 .
\eeq
This is  the desired Fokker-Planck equation, provided that 
the initial distribution $\bar{f}_0(\bfu , t_0)$ is 
identified with the averaged distribution function $P(\bfu ,t)$.
This means that the initial distribution function
at an arbitrary time $t=t_0$  before averaging  must coincide
with  the averaged distribution to be determined.
The initial value may be considered as the integral constant
 in the unperturbed equation, which would move slowly being governed
 by the RG equation.
In other words, the averaging is automatically made by the
RG method.

\end{document}

%% file: rg_fig.tex
\setlength{\unitlength}{0.240900pt}
\ifx\plotpoint\undefined\newsavebox{\plotpoint}\fi
\sbox{\plotpoint}{\rule[-0.200pt]{0.400pt}{0.400pt}}%
\begin{picture}(1500,900)(0,0)
\font\gnuplot=cmr10 at 10pt
\gnuplot
\sbox{\plotpoint}{\rule[-0.200pt]{0.400pt}{0.400pt}}%
\put(60.0,61.0){\rule[-0.200pt]{332.201pt}{0.400pt}}
\put(1439.0,61.0){\rule[-0.200pt]{0.400pt}{172.484pt}}
\put(60.0,777.0){\rule[-0.200pt]{332.201pt}{0.400pt}}
\put(749,21){\makebox(0,0){$t$ }}
\put(749,839){\makebox(0,0){Distribution functions}}
\put(511,663){\makebox(0,0)[l]{$f(\bfu, t)$}}
\put(556,582){\makebox(0,0)[l]{$P(\bfu, t_0)=\tilde{f}(\bfu, t=t_0; t_0)$}}
\put(474,23){\makebox(0,0)[l]{$t_0$}}
\put(60.0,61.0){\rule[-0.200pt]{0.400pt}{172.484pt}}
\multiput(508.92,632.37)(-0.498,-1.275){97}{\rule{0.120pt}{1.116pt}}
\multiput(509.17,634.68)(-50.000,-124.684){2}{\rule{0.400pt}{0.558pt}}
\put(460,510){\vector(-1,-3){0}}
\multiput(552.92,566.47)(-0.499,-0.638){157}{\rule{0.120pt}{0.610pt}}
\multiput(553.17,567.73)(-80.000,-100.734){2}{\rule{0.400pt}{0.305pt}}
\put(474,467){\vector(-3,-4){0}}
\put(474,467){\vector(0,-1){406}}
\put(60,700){\usebox{\plotpoint}}
\multiput(60.00,698.92)(0.704,-0.491){17}{\rule{0.660pt}{0.118pt}}
\multiput(60.00,699.17)(12.630,-10.000){2}{\rule{0.330pt}{0.400pt}}
\multiput(74.00,688.92)(0.637,-0.492){19}{\rule{0.609pt}{0.118pt}}
\multiput(74.00,689.17)(12.736,-11.000){2}{\rule{0.305pt}{0.400pt}}
\multiput(88.00,677.92)(0.704,-0.491){17}{\rule{0.660pt}{0.118pt}}
\multiput(88.00,678.17)(12.630,-10.000){2}{\rule{0.330pt}{0.400pt}}
\multiput(102.00,667.92)(0.704,-0.491){17}{\rule{0.660pt}{0.118pt}}
\multiput(102.00,668.17)(12.630,-10.000){2}{\rule{0.330pt}{0.400pt}}
\multiput(116.00,657.92)(0.704,-0.491){17}{\rule{0.660pt}{0.118pt}}
\multiput(116.00,658.17)(12.630,-10.000){2}{\rule{0.330pt}{0.400pt}}
\multiput(130.00,647.93)(0.786,-0.489){15}{\rule{0.722pt}{0.118pt}}
\multiput(130.00,648.17)(12.501,-9.000){2}{\rule{0.361pt}{0.400pt}}
\multiput(144.00,638.92)(0.704,-0.491){17}{\rule{0.660pt}{0.118pt}}
\multiput(144.00,639.17)(12.630,-10.000){2}{\rule{0.330pt}{0.400pt}}
\multiput(158.00,628.93)(0.728,-0.489){15}{\rule{0.678pt}{0.118pt}}
\multiput(158.00,629.17)(11.593,-9.000){2}{\rule{0.339pt}{0.400pt}}
\multiput(171.00,619.93)(0.786,-0.489){15}{\rule{0.722pt}{0.118pt}}
\multiput(171.00,620.17)(12.501,-9.000){2}{\rule{0.361pt}{0.400pt}}
\multiput(185.00,610.93)(0.890,-0.488){13}{\rule{0.800pt}{0.117pt}}
\multiput(185.00,611.17)(12.340,-8.000){2}{\rule{0.400pt}{0.400pt}}
\multiput(199.00,602.93)(0.786,-0.489){15}{\rule{0.722pt}{0.118pt}}
\multiput(199.00,603.17)(12.501,-9.000){2}{\rule{0.361pt}{0.400pt}}
\multiput(213.00,593.93)(0.890,-0.488){13}{\rule{0.800pt}{0.117pt}}
\multiput(213.00,594.17)(12.340,-8.000){2}{\rule{0.400pt}{0.400pt}}
\multiput(227.00,585.93)(0.890,-0.488){13}{\rule{0.800pt}{0.117pt}}
\multiput(227.00,586.17)(12.340,-8.000){2}{\rule{0.400pt}{0.400pt}}
\multiput(241.00,577.93)(0.890,-0.488){13}{\rule{0.800pt}{0.117pt}}
\multiput(241.00,578.17)(12.340,-8.000){2}{\rule{0.400pt}{0.400pt}}
\multiput(255.00,569.93)(0.890,-0.488){13}{\rule{0.800pt}{0.117pt}}
\multiput(255.00,570.17)(12.340,-8.000){2}{\rule{0.400pt}{0.400pt}}
\multiput(269.00,561.93)(0.890,-0.488){13}{\rule{0.800pt}{0.117pt}}
\multiput(269.00,562.17)(12.340,-8.000){2}{\rule{0.400pt}{0.400pt}}
\multiput(283.00,553.93)(1.026,-0.485){11}{\rule{0.900pt}{0.117pt}}
\multiput(283.00,554.17)(12.132,-7.000){2}{\rule{0.450pt}{0.400pt}}
\multiput(297.00,546.93)(1.026,-0.485){11}{\rule{0.900pt}{0.117pt}}
\multiput(297.00,547.17)(12.132,-7.000){2}{\rule{0.450pt}{0.400pt}}
\multiput(311.00,539.93)(1.026,-0.485){11}{\rule{0.900pt}{0.117pt}}
\multiput(311.00,540.17)(12.132,-7.000){2}{\rule{0.450pt}{0.400pt}}
\multiput(325.00,532.93)(1.026,-0.485){11}{\rule{0.900pt}{0.117pt}}
\multiput(325.00,533.17)(12.132,-7.000){2}{\rule{0.450pt}{0.400pt}}
\multiput(339.00,525.93)(1.026,-0.485){11}{\rule{0.900pt}{0.117pt}}
\multiput(339.00,526.17)(12.132,-7.000){2}{\rule{0.450pt}{0.400pt}}
\multiput(353.00,518.93)(0.950,-0.485){11}{\rule{0.843pt}{0.117pt}}
\multiput(353.00,519.17)(11.251,-7.000){2}{\rule{0.421pt}{0.400pt}}
\multiput(366.00,511.93)(1.214,-0.482){9}{\rule{1.033pt}{0.116pt}}
\multiput(366.00,512.17)(11.855,-6.000){2}{\rule{0.517pt}{0.400pt}}
\multiput(380.00,505.93)(1.026,-0.485){11}{\rule{0.900pt}{0.117pt}}
\multiput(380.00,506.17)(12.132,-7.000){2}{\rule{0.450pt}{0.400pt}}
\multiput(394.00,498.93)(1.214,-0.482){9}{\rule{1.033pt}{0.116pt}}
\multiput(394.00,499.17)(11.855,-6.000){2}{\rule{0.517pt}{0.400pt}}
\multiput(408.00,492.93)(1.214,-0.482){9}{\rule{1.033pt}{0.116pt}}
\multiput(408.00,493.17)(11.855,-6.000){2}{\rule{0.517pt}{0.400pt}}
\multiput(422.00,486.93)(1.214,-0.482){9}{\rule{1.033pt}{0.116pt}}
\multiput(422.00,487.17)(11.855,-6.000){2}{\rule{0.517pt}{0.400pt}}
\multiput(436.00,480.93)(1.214,-0.482){9}{\rule{1.033pt}{0.116pt}}
\multiput(436.00,481.17)(11.855,-6.000){2}{\rule{0.517pt}{0.400pt}}
\multiput(450.00,474.93)(1.214,-0.482){9}{\rule{1.033pt}{0.116pt}}
\multiput(450.00,475.17)(11.855,-6.000){2}{\rule{0.517pt}{0.400pt}}
\multiput(464.00,468.93)(1.489,-0.477){7}{\rule{1.220pt}{0.115pt}}
\multiput(464.00,469.17)(11.468,-5.000){2}{\rule{0.610pt}{0.400pt}}
\multiput(478.00,463.93)(1.214,-0.482){9}{\rule{1.033pt}{0.116pt}}
\multiput(478.00,464.17)(11.855,-6.000){2}{\rule{0.517pt}{0.400pt}}
\multiput(492.00,457.93)(1.489,-0.477){7}{\rule{1.220pt}{0.115pt}}
\multiput(492.00,458.17)(11.468,-5.000){2}{\rule{0.610pt}{0.400pt}}
\multiput(506.00,452.93)(1.489,-0.477){7}{\rule{1.220pt}{0.115pt}}
\multiput(506.00,453.17)(11.468,-5.000){2}{\rule{0.610pt}{0.400pt}}
\multiput(520.00,447.93)(1.489,-0.477){7}{\rule{1.220pt}{0.115pt}}
\multiput(520.00,448.17)(11.468,-5.000){2}{\rule{0.610pt}{0.400pt}}
\multiput(534.00,442.93)(1.489,-0.477){7}{\rule{1.220pt}{0.115pt}}
\multiput(534.00,443.17)(11.468,-5.000){2}{\rule{0.610pt}{0.400pt}}
\multiput(548.00,437.93)(1.378,-0.477){7}{\rule{1.140pt}{0.115pt}}
\multiput(548.00,438.17)(10.634,-5.000){2}{\rule{0.570pt}{0.400pt}}
\multiput(561.00,432.93)(1.489,-0.477){7}{\rule{1.220pt}{0.115pt}}
\multiput(561.00,433.17)(11.468,-5.000){2}{\rule{0.610pt}{0.400pt}}
\multiput(575.00,427.93)(1.489,-0.477){7}{\rule{1.220pt}{0.115pt}}
\multiput(575.00,428.17)(11.468,-5.000){2}{\rule{0.610pt}{0.400pt}}
\multiput(589.00,422.94)(1.943,-0.468){5}{\rule{1.500pt}{0.113pt}}
\multiput(589.00,423.17)(10.887,-4.000){2}{\rule{0.750pt}{0.400pt}}
\multiput(603.00,418.93)(1.489,-0.477){7}{\rule{1.220pt}{0.115pt}}
\multiput(603.00,419.17)(11.468,-5.000){2}{\rule{0.610pt}{0.400pt}}
\multiput(617.00,413.94)(1.943,-0.468){5}{\rule{1.500pt}{0.113pt}}
\multiput(617.00,414.17)(10.887,-4.000){2}{\rule{0.750pt}{0.400pt}}
\multiput(631.00,409.93)(1.489,-0.477){7}{\rule{1.220pt}{0.115pt}}
\multiput(631.00,410.17)(11.468,-5.000){2}{\rule{0.610pt}{0.400pt}}
\multiput(645.00,404.94)(1.943,-0.468){5}{\rule{1.500pt}{0.113pt}}
\multiput(645.00,405.17)(10.887,-4.000){2}{\rule{0.750pt}{0.400pt}}
\multiput(659.00,400.94)(1.943,-0.468){5}{\rule{1.500pt}{0.113pt}}
\multiput(659.00,401.17)(10.887,-4.000){2}{\rule{0.750pt}{0.400pt}}
\multiput(673.00,396.94)(1.943,-0.468){5}{\rule{1.500pt}{0.113pt}}
\multiput(673.00,397.17)(10.887,-4.000){2}{\rule{0.750pt}{0.400pt}}
\multiput(687.00,392.94)(1.943,-0.468){5}{\rule{1.500pt}{0.113pt}}
\multiput(687.00,393.17)(10.887,-4.000){2}{\rule{0.750pt}{0.400pt}}
\multiput(701.00,388.94)(1.943,-0.468){5}{\rule{1.500pt}{0.113pt}}
\multiput(701.00,389.17)(10.887,-4.000){2}{\rule{0.750pt}{0.400pt}}
\multiput(715.00,384.94)(1.943,-0.468){5}{\rule{1.500pt}{0.113pt}}
\multiput(715.00,385.17)(10.887,-4.000){2}{\rule{0.750pt}{0.400pt}}
\multiput(729.00,380.95)(2.918,-0.447){3}{\rule{1.967pt}{0.108pt}}
\multiput(729.00,381.17)(9.918,-3.000){2}{\rule{0.983pt}{0.400pt}}
\multiput(743.00,377.94)(1.797,-0.468){5}{\rule{1.400pt}{0.113pt}}
\multiput(743.00,378.17)(10.094,-4.000){2}{\rule{0.700pt}{0.400pt}}
\multiput(756.00,373.94)(1.943,-0.468){5}{\rule{1.500pt}{0.113pt}}
\multiput(756.00,374.17)(10.887,-4.000){2}{\rule{0.750pt}{0.400pt}}
\multiput(770.00,369.95)(2.918,-0.447){3}{\rule{1.967pt}{0.108pt}}
\multiput(770.00,370.17)(9.918,-3.000){2}{\rule{0.983pt}{0.400pt}}
\multiput(784.00,366.94)(1.943,-0.468){5}{\rule{1.500pt}{0.113pt}}
\multiput(784.00,367.17)(10.887,-4.000){2}{\rule{0.750pt}{0.400pt}}
\multiput(798.00,362.95)(2.918,-0.447){3}{\rule{1.967pt}{0.108pt}}
\multiput(798.00,363.17)(9.918,-3.000){2}{\rule{0.983pt}{0.400pt}}
\multiput(812.00,359.95)(2.918,-0.447){3}{\rule{1.967pt}{0.108pt}}
\multiput(812.00,360.17)(9.918,-3.000){2}{\rule{0.983pt}{0.400pt}}
\multiput(826.00,356.94)(1.943,-0.468){5}{\rule{1.500pt}{0.113pt}}
\multiput(826.00,357.17)(10.887,-4.000){2}{\rule{0.750pt}{0.400pt}}
\multiput(840.00,352.95)(2.918,-0.447){3}{\rule{1.967pt}{0.108pt}}
\multiput(840.00,353.17)(9.918,-3.000){2}{\rule{0.983pt}{0.400pt}}
\multiput(854.00,349.95)(2.918,-0.447){3}{\rule{1.967pt}{0.108pt}}
\multiput(854.00,350.17)(9.918,-3.000){2}{\rule{0.983pt}{0.400pt}}
\multiput(868.00,346.95)(2.918,-0.447){3}{\rule{1.967pt}{0.108pt}}
\multiput(868.00,347.17)(9.918,-3.000){2}{\rule{0.983pt}{0.400pt}}
\multiput(882.00,343.95)(2.918,-0.447){3}{\rule{1.967pt}{0.108pt}}
\multiput(882.00,344.17)(9.918,-3.000){2}{\rule{0.983pt}{0.400pt}}
\multiput(896.00,340.95)(2.918,-0.447){3}{\rule{1.967pt}{0.108pt}}
\multiput(896.00,341.17)(9.918,-3.000){2}{\rule{0.983pt}{0.400pt}}
\multiput(910.00,337.95)(2.918,-0.447){3}{\rule{1.967pt}{0.108pt}}
\multiput(910.00,338.17)(9.918,-3.000){2}{\rule{0.983pt}{0.400pt}}
\multiput(924.00,334.95)(2.918,-0.447){3}{\rule{1.967pt}{0.108pt}}
\multiput(924.00,335.17)(9.918,-3.000){2}{\rule{0.983pt}{0.400pt}}
\multiput(938.00,331.95)(2.695,-0.447){3}{\rule{1.833pt}{0.108pt}}
\multiput(938.00,332.17)(9.195,-3.000){2}{\rule{0.917pt}{0.400pt}}
\put(951,328.17){\rule{2.900pt}{0.400pt}}
\multiput(951.00,329.17)(7.981,-2.000){2}{\rule{1.450pt}{0.400pt}}
\multiput(965.00,326.95)(2.918,-0.447){3}{\rule{1.967pt}{0.108pt}}
\multiput(965.00,327.17)(9.918,-3.000){2}{\rule{0.983pt}{0.400pt}}
\multiput(979.00,323.95)(2.918,-0.447){3}{\rule{1.967pt}{0.108pt}}
\multiput(979.00,324.17)(9.918,-3.000){2}{\rule{0.983pt}{0.400pt}}
\put(993,320.17){\rule{2.900pt}{0.400pt}}
\multiput(993.00,321.17)(7.981,-2.000){2}{\rule{1.450pt}{0.400pt}}
\multiput(1007.00,318.95)(2.918,-0.447){3}{\rule{1.967pt}{0.108pt}}
\multiput(1007.00,319.17)(9.918,-3.000){2}{\rule{0.983pt}{0.400pt}}
\put(1021,315.17){\rule{2.900pt}{0.400pt}}
\multiput(1021.00,316.17)(7.981,-2.000){2}{\rule{1.450pt}{0.400pt}}
\multiput(1035.00,313.95)(2.918,-0.447){3}{\rule{1.967pt}{0.108pt}}
\multiput(1035.00,314.17)(9.918,-3.000){2}{\rule{0.983pt}{0.400pt}}
\put(1049,310.17){\rule{2.900pt}{0.400pt}}
\multiput(1049.00,311.17)(7.981,-2.000){2}{\rule{1.450pt}{0.400pt}}
\put(1063,308.17){\rule{2.900pt}{0.400pt}}
\multiput(1063.00,309.17)(7.981,-2.000){2}{\rule{1.450pt}{0.400pt}}
\multiput(1077.00,306.95)(2.918,-0.447){3}{\rule{1.967pt}{0.108pt}}
\multiput(1077.00,307.17)(9.918,-3.000){2}{\rule{0.983pt}{0.400pt}}
\put(1091,303.17){\rule{2.900pt}{0.400pt}}
\multiput(1091.00,304.17)(7.981,-2.000){2}{\rule{1.450pt}{0.400pt}}
\put(1105,301.17){\rule{2.900pt}{0.400pt}}
\multiput(1105.00,302.17)(7.981,-2.000){2}{\rule{1.450pt}{0.400pt}}
\put(1119,299.17){\rule{2.900pt}{0.400pt}}
\multiput(1119.00,300.17)(7.981,-2.000){2}{\rule{1.450pt}{0.400pt}}
\multiput(1133.00,297.95)(2.695,-0.447){3}{\rule{1.833pt}{0.108pt}}
\multiput(1133.00,298.17)(9.195,-3.000){2}{\rule{0.917pt}{0.400pt}}
\put(1146,294.17){\rule{2.900pt}{0.400pt}}
\multiput(1146.00,295.17)(7.981,-2.000){2}{\rule{1.450pt}{0.400pt}}
\put(1160,292.17){\rule{2.900pt}{0.400pt}}
\multiput(1160.00,293.17)(7.981,-2.000){2}{\rule{1.450pt}{0.400pt}}
\put(1174,290.17){\rule{2.900pt}{0.400pt}}
\multiput(1174.00,291.17)(7.981,-2.000){2}{\rule{1.450pt}{0.400pt}}
\put(1188,288.17){\rule{2.900pt}{0.400pt}}
\multiput(1188.00,289.17)(7.981,-2.000){2}{\rule{1.450pt}{0.400pt}}
\put(1202,286.17){\rule{2.900pt}{0.400pt}}
\multiput(1202.00,287.17)(7.981,-2.000){2}{\rule{1.450pt}{0.400pt}}
\put(1216,284.17){\rule{2.900pt}{0.400pt}}
\multiput(1216.00,285.17)(7.981,-2.000){2}{\rule{1.450pt}{0.400pt}}
\put(1230,282.17){\rule{2.900pt}{0.400pt}}
\multiput(1230.00,283.17)(7.981,-2.000){2}{\rule{1.450pt}{0.400pt}}
\put(1244,280.17){\rule{2.900pt}{0.400pt}}
\multiput(1244.00,281.17)(7.981,-2.000){2}{\rule{1.450pt}{0.400pt}}
\put(1258,278.67){\rule{3.373pt}{0.400pt}}
\multiput(1258.00,279.17)(7.000,-1.000){2}{\rule{1.686pt}{0.400pt}}
\put(1272,277.17){\rule{2.900pt}{0.400pt}}
\multiput(1272.00,278.17)(7.981,-2.000){2}{\rule{1.450pt}{0.400pt}}
\put(1286,275.17){\rule{2.900pt}{0.400pt}}
\multiput(1286.00,276.17)(7.981,-2.000){2}{\rule{1.450pt}{0.400pt}}
\put(1300,273.17){\rule{2.900pt}{0.400pt}}
\multiput(1300.00,274.17)(7.981,-2.000){2}{\rule{1.450pt}{0.400pt}}
\put(1314,271.17){\rule{2.900pt}{0.400pt}}
\multiput(1314.00,272.17)(7.981,-2.000){2}{\rule{1.450pt}{0.400pt}}
\put(1328,269.67){\rule{3.132pt}{0.400pt}}
\multiput(1328.00,270.17)(6.500,-1.000){2}{\rule{1.566pt}{0.400pt}}
\put(1341,268.17){\rule{2.900pt}{0.400pt}}
\multiput(1341.00,269.17)(7.981,-2.000){2}{\rule{1.450pt}{0.400pt}}
\put(1355,266.17){\rule{2.900pt}{0.400pt}}
\multiput(1355.00,267.17)(7.981,-2.000){2}{\rule{1.450pt}{0.400pt}}
\put(1369,264.67){\rule{3.373pt}{0.400pt}}
\multiput(1369.00,265.17)(7.000,-1.000){2}{\rule{1.686pt}{0.400pt}}
\put(1383,263.17){\rule{2.900pt}{0.400pt}}
\multiput(1383.00,264.17)(7.981,-2.000){2}{\rule{1.450pt}{0.400pt}}
\put(1397,261.67){\rule{3.373pt}{0.400pt}}
\multiput(1397.00,262.17)(7.000,-1.000){2}{\rule{1.686pt}{0.400pt}}
\put(1411,260.17){\rule{2.900pt}{0.400pt}}
\multiput(1411.00,261.17)(7.981,-2.000){2}{\rule{1.450pt}{0.400pt}}
\put(1425,258.67){\rule{3.373pt}{0.400pt}}
\multiput(1425.00,259.17)(7.000,-1.000){2}{\rule{1.686pt}{0.400pt}}
\put(60,700){\usebox{\plotpoint}}
\multiput(60.58,700.00)(0.494,2.076){25}{\rule{0.119pt}{1.729pt}}
\multiput(59.17,700.00)(14.000,53.412){2}{\rule{0.400pt}{0.864pt}}
\multiput(74.58,753.98)(0.494,-0.791){25}{\rule{0.119pt}{0.729pt}}
\multiput(73.17,755.49)(14.000,-20.488){2}{\rule{0.400pt}{0.364pt}}
\multiput(88.58,727.94)(0.494,-2.039){25}{\rule{0.119pt}{1.700pt}}
\multiput(87.17,731.47)(14.000,-52.472){2}{\rule{0.400pt}{0.850pt}}
\multiput(102.58,675.98)(0.494,-0.791){25}{\rule{0.119pt}{0.729pt}}
\multiput(101.17,677.49)(14.000,-20.488){2}{\rule{0.400pt}{0.364pt}}
\multiput(116.00,655.92)(0.637,-0.492){19}{\rule{0.609pt}{0.118pt}}
\multiput(116.00,656.17)(12.736,-11.000){2}{\rule{0.305pt}{0.400pt}}
\multiput(130.58,640.84)(0.494,-1.452){25}{\rule{0.119pt}{1.243pt}}
\multiput(129.17,643.42)(14.000,-37.420){2}{\rule{0.400pt}{0.621pt}}
\multiput(144.58,601.79)(0.494,-1.158){25}{\rule{0.119pt}{1.014pt}}
\multiput(143.17,603.89)(14.000,-29.895){2}{\rule{0.400pt}{0.507pt}}
\multiput(158.58,574.00)(0.493,0.655){23}{\rule{0.119pt}{0.623pt}}
\multiput(157.17,574.00)(13.000,15.707){2}{\rule{0.400pt}{0.312pt}}
\multiput(171.58,591.00)(0.494,1.488){25}{\rule{0.119pt}{1.271pt}}
\multiput(170.17,591.00)(14.000,38.361){2}{\rule{0.400pt}{0.636pt}}
\multiput(185.58,632.00)(0.494,0.827){25}{\rule{0.119pt}{0.757pt}}
\multiput(184.17,632.00)(14.000,21.429){2}{\rule{0.400pt}{0.379pt}}
\multiput(199.00,653.93)(1.214,-0.482){9}{\rule{1.033pt}{0.116pt}}
\multiput(199.00,654.17)(11.855,-6.000){2}{\rule{0.517pt}{0.400pt}}
\multiput(213.58,643.96)(0.494,-1.415){25}{\rule{0.119pt}{1.214pt}}
\multiput(212.17,646.48)(14.000,-36.480){2}{\rule{0.400pt}{0.607pt}}
\multiput(227.58,603.06)(0.494,-2.003){25}{\rule{0.119pt}{1.671pt}}
\multiput(226.17,606.53)(14.000,-51.531){2}{\rule{0.400pt}{0.836pt}}
\multiput(241.58,551.26)(0.494,-1.011){25}{\rule{0.119pt}{0.900pt}}
\multiput(240.17,553.13)(14.000,-26.132){2}{\rule{0.400pt}{0.450pt}}
\multiput(255.00,527.58)(0.637,0.492){19}{\rule{0.609pt}{0.118pt}}
\multiput(255.00,526.17)(12.736,11.000){2}{\rule{0.305pt}{0.400pt}}
\multiput(269.00,538.60)(1.943,0.468){5}{\rule{1.500pt}{0.113pt}}
\multiput(269.00,537.17)(10.887,4.000){2}{\rule{0.750pt}{0.400pt}}
\multiput(283.58,539.81)(0.494,-0.534){25}{\rule{0.119pt}{0.529pt}}
\multiput(282.17,540.90)(14.000,-13.903){2}{\rule{0.400pt}{0.264pt}}
\multiput(297.58,527.00)(0.494,0.644){25}{\rule{0.119pt}{0.614pt}}
\multiput(296.17,527.00)(14.000,16.725){2}{\rule{0.400pt}{0.307pt}}
\multiput(311.58,545.00)(0.494,1.819){25}{\rule{0.119pt}{1.529pt}}
\multiput(310.17,545.00)(14.000,46.827){2}{\rule{0.400pt}{0.764pt}}
\multiput(325.00,593.94)(1.943,-0.468){5}{\rule{1.500pt}{0.113pt}}
\multiput(325.00,594.17)(10.887,-4.000){2}{\rule{0.750pt}{0.400pt}}
\multiput(339.58,581.45)(0.494,-2.811){25}{\rule{0.119pt}{2.300pt}}
\multiput(338.17,586.23)(14.000,-72.226){2}{\rule{0.400pt}{1.150pt}}
\multiput(353.58,506.56)(0.493,-2.162){23}{\rule{0.119pt}{1.792pt}}
\multiput(352.17,510.28)(13.000,-51.280){2}{\rule{0.400pt}{0.896pt}}
\multiput(366.58,459.00)(0.494,0.570){25}{\rule{0.119pt}{0.557pt}}
\multiput(365.17,459.00)(14.000,14.844){2}{\rule{0.400pt}{0.279pt}}
\multiput(380.58,475.00)(0.494,0.754){25}{\rule{0.119pt}{0.700pt}}
\multiput(379.17,475.00)(14.000,19.547){2}{\rule{0.400pt}{0.350pt}}
\multiput(394.58,493.69)(0.494,-0.570){25}{\rule{0.119pt}{0.557pt}}
\multiput(393.17,494.84)(14.000,-14.844){2}{\rule{0.400pt}{0.279pt}}
\multiput(408.00,478.95)(2.918,-0.447){3}{\rule{1.967pt}{0.108pt}}
\multiput(408.00,479.17)(9.918,-3.000){2}{\rule{0.983pt}{0.400pt}}
\multiput(422.58,477.00)(0.494,1.158){25}{\rule{0.119pt}{1.014pt}}
\multiput(421.17,477.00)(14.000,29.895){2}{\rule{0.400pt}{0.507pt}}
\multiput(436.58,509.00)(0.494,0.607){25}{\rule{0.119pt}{0.586pt}}
\multiput(435.17,509.00)(14.000,15.784){2}{\rule{0.400pt}{0.293pt}}
\multiput(450.58,522.26)(0.494,-1.011){25}{\rule{0.119pt}{0.900pt}}
\multiput(449.17,524.13)(14.000,-26.132){2}{\rule{0.400pt}{0.450pt}}
\multiput(464.58,492.37)(0.494,-1.599){25}{\rule{0.119pt}{1.357pt}}
\multiput(463.17,495.18)(14.000,-41.183){2}{\rule{0.400pt}{0.679pt}}
\multiput(478.58,450.15)(0.494,-1.048){25}{\rule{0.119pt}{0.929pt}}
\multiput(477.17,452.07)(14.000,-27.073){2}{\rule{0.400pt}{0.464pt}}
\multiput(492.00,423.93)(0.786,-0.489){15}{\rule{0.722pt}{0.118pt}}
\multiput(492.00,424.17)(12.501,-9.000){2}{\rule{0.361pt}{0.400pt}}
\multiput(506.58,416.00)(0.494,0.644){25}{\rule{0.119pt}{0.614pt}}
\multiput(505.17,416.00)(14.000,16.725){2}{\rule{0.400pt}{0.307pt}}
\multiput(520.58,434.00)(0.494,0.938){25}{\rule{0.119pt}{0.843pt}}
\multiput(519.17,434.00)(14.000,24.251){2}{\rule{0.400pt}{0.421pt}}
\put(534,458.17){\rule{2.900pt}{0.400pt}}
\multiput(534.00,459.17)(7.981,-2.000){2}{\rule{1.450pt}{0.400pt}}
\multiput(548.58,454.90)(0.493,-0.814){23}{\rule{0.119pt}{0.746pt}}
\multiput(547.17,456.45)(13.000,-19.451){2}{\rule{0.400pt}{0.373pt}}
\multiput(561.00,437.59)(1.489,0.477){7}{\rule{1.220pt}{0.115pt}}
\multiput(561.00,436.17)(11.468,5.000){2}{\rule{0.610pt}{0.400pt}}
\multiput(575.58,442.00)(0.494,0.534){25}{\rule{0.119pt}{0.529pt}}
\multiput(574.17,442.00)(14.000,13.903){2}{\rule{0.400pt}{0.264pt}}
\multiput(589.58,452.55)(0.494,-1.231){25}{\rule{0.119pt}{1.071pt}}
\multiput(588.17,454.78)(14.000,-31.776){2}{\rule{0.400pt}{0.536pt}}
\multiput(603.58,415.82)(0.494,-2.076){25}{\rule{0.119pt}{1.729pt}}
\multiput(602.17,419.41)(14.000,-53.412){2}{\rule{0.400pt}{0.864pt}}
\multiput(617.00,366.60)(1.943,0.468){5}{\rule{1.500pt}{0.113pt}}
\multiput(617.00,365.17)(10.887,4.000){2}{\rule{0.750pt}{0.400pt}}
\multiput(631.58,370.00)(0.494,1.929){25}{\rule{0.119pt}{1.614pt}}
\multiput(630.17,370.00)(14.000,49.649){2}{\rule{0.400pt}{0.807pt}}
\multiput(645.00,423.58)(0.637,0.492){19}{\rule{0.609pt}{0.118pt}}
\multiput(645.00,422.17)(12.736,11.000){2}{\rule{0.305pt}{0.400pt}}
\multiput(659.58,428.96)(0.494,-1.415){25}{\rule{0.119pt}{1.214pt}}
\multiput(658.17,431.48)(14.000,-36.480){2}{\rule{0.400pt}{0.607pt}}
\multiput(673.00,393.92)(0.582,-0.492){21}{\rule{0.567pt}{0.119pt}}
\multiput(673.00,394.17)(12.824,-12.000){2}{\rule{0.283pt}{0.400pt}}
\multiput(687.58,383.00)(0.494,0.901){25}{\rule{0.119pt}{0.814pt}}
\multiput(686.17,383.00)(14.000,23.310){2}{\rule{0.400pt}{0.407pt}}
\multiput(715.58,403.20)(0.494,-1.342){25}{\rule{0.119pt}{1.157pt}}
\multiput(714.17,405.60)(14.000,-34.598){2}{\rule{0.400pt}{0.579pt}}
\multiput(729.58,368.33)(0.494,-0.680){25}{\rule{0.119pt}{0.643pt}}
\multiput(728.17,369.67)(14.000,-17.666){2}{\rule{0.400pt}{0.321pt}}
\multiput(743.00,352.58)(0.590,0.492){19}{\rule{0.573pt}{0.118pt}}
\multiput(743.00,351.17)(11.811,11.000){2}{\rule{0.286pt}{0.400pt}}
\multiput(756.00,363.58)(0.497,0.494){25}{\rule{0.500pt}{0.119pt}}
\multiput(756.00,362.17)(12.962,14.000){2}{\rule{0.250pt}{0.400pt}}
\multiput(770.00,377.61)(2.918,0.447){3}{\rule{1.967pt}{0.108pt}}
\multiput(770.00,376.17)(9.918,3.000){2}{\rule{0.983pt}{0.400pt}}
\multiput(784.00,378.93)(1.026,-0.485){11}{\rule{0.900pt}{0.117pt}}
\multiput(784.00,379.17)(12.132,-7.000){2}{\rule{0.450pt}{0.400pt}}
\multiput(798.58,370.45)(0.494,-0.644){25}{\rule{0.119pt}{0.614pt}}
\multiput(797.17,371.73)(14.000,-16.725){2}{\rule{0.400pt}{0.307pt}}
\multiput(812.00,353.92)(0.637,-0.492){19}{\rule{0.609pt}{0.118pt}}
\multiput(812.00,354.17)(12.736,-11.000){2}{\rule{0.305pt}{0.400pt}}
\multiput(826.58,344.00)(0.494,0.680){25}{\rule{0.119pt}{0.643pt}}
\multiput(825.17,344.00)(14.000,17.666){2}{\rule{0.400pt}{0.321pt}}
\multiput(840.00,363.58)(0.497,0.494){25}{\rule{0.500pt}{0.119pt}}
\multiput(840.00,362.17)(12.962,14.000){2}{\rule{0.250pt}{0.400pt}}
\multiput(854.58,372.91)(0.494,-1.121){25}{\rule{0.119pt}{0.986pt}}
\multiput(853.17,374.95)(14.000,-28.954){2}{\rule{0.400pt}{0.493pt}}
\multiput(868.58,341.79)(0.494,-1.158){25}{\rule{0.119pt}{1.014pt}}
\multiput(867.17,343.89)(14.000,-29.895){2}{\rule{0.400pt}{0.507pt}}
\multiput(882.58,314.00)(0.494,0.754){25}{\rule{0.119pt}{0.700pt}}
\multiput(881.17,314.00)(14.000,19.547){2}{\rule{0.400pt}{0.350pt}}
\multiput(896.58,335.00)(0.494,1.121){25}{\rule{0.119pt}{0.986pt}}
\multiput(895.17,335.00)(14.000,28.954){2}{\rule{0.400pt}{0.493pt}}
\multiput(910.58,362.50)(0.494,-0.938){25}{\rule{0.119pt}{0.843pt}}
\multiput(909.17,364.25)(14.000,-24.251){2}{\rule{0.400pt}{0.421pt}}
\multiput(924.58,334.60)(0.494,-1.525){25}{\rule{0.119pt}{1.300pt}}
\multiput(923.17,337.30)(14.000,-39.302){2}{\rule{0.400pt}{0.650pt}}
\multiput(938.58,298.00)(0.493,0.774){23}{\rule{0.119pt}{0.715pt}}
\multiput(937.17,298.00)(13.000,18.515){2}{\rule{0.400pt}{0.358pt}}
\multiput(951.58,318.00)(0.494,1.672){25}{\rule{0.119pt}{1.414pt}}
\multiput(950.17,318.00)(14.000,43.065){2}{\rule{0.400pt}{0.707pt}}
\multiput(965.00,362.93)(0.890,-0.488){13}{\rule{0.800pt}{0.117pt}}
\multiput(965.00,363.17)(12.340,-8.000){2}{\rule{0.400pt}{0.400pt}}
\multiput(979.58,350.37)(0.494,-1.599){25}{\rule{0.119pt}{1.357pt}}
\multiput(978.17,353.18)(14.000,-41.183){2}{\rule{0.400pt}{0.679pt}}
\multiput(993.00,310.92)(0.536,-0.493){23}{\rule{0.531pt}{0.119pt}}
\multiput(993.00,311.17)(12.898,-13.000){2}{\rule{0.265pt}{0.400pt}}
\multiput(1007.00,299.58)(0.497,0.494){25}{\rule{0.500pt}{0.119pt}}
\multiput(1007.00,298.17)(12.962,14.000){2}{\rule{0.250pt}{0.400pt}}
\put(1021,311.67){\rule{3.373pt}{0.400pt}}
\multiput(1021.00,312.17)(7.000,-1.000){2}{\rule{1.686pt}{0.400pt}}
\multiput(1035.00,310.92)(0.536,-0.493){23}{\rule{0.531pt}{0.119pt}}
\multiput(1035.00,311.17)(12.898,-13.000){2}{\rule{0.265pt}{0.400pt}}
\put(1049,298.67){\rule{3.373pt}{0.400pt}}
\multiput(1049.00,298.17)(7.000,1.000){2}{\rule{1.686pt}{0.400pt}}
\multiput(1063.00,300.58)(0.536,0.493){23}{\rule{0.531pt}{0.119pt}}
\multiput(1063.00,299.17)(12.898,13.000){2}{\rule{0.265pt}{0.400pt}}
\multiput(1077.00,313.58)(0.497,0.494){25}{\rule{0.500pt}{0.119pt}}
\multiput(1077.00,312.17)(12.962,14.000){2}{\rule{0.250pt}{0.400pt}}
\multiput(1091.00,327.58)(0.704,0.491){17}{\rule{0.660pt}{0.118pt}}
\multiput(1091.00,326.17)(12.630,10.000){2}{\rule{0.330pt}{0.400pt}}
\multiput(1105.58,334.45)(0.494,-0.644){25}{\rule{0.119pt}{0.614pt}}
\multiput(1104.17,335.73)(14.000,-16.725){2}{\rule{0.400pt}{0.307pt}}
\multiput(1119.58,312.89)(0.494,-1.745){25}{\rule{0.119pt}{1.471pt}}
\multiput(1118.17,315.95)(14.000,-44.946){2}{\rule{0.400pt}{0.736pt}}
\multiput(1133.58,267.77)(0.493,-0.853){23}{\rule{0.119pt}{0.777pt}}
\multiput(1132.17,269.39)(13.000,-20.387){2}{\rule{0.400pt}{0.388pt}}
\multiput(1146.58,249.00)(0.494,1.158){25}{\rule{0.119pt}{1.014pt}}
\multiput(1145.17,249.00)(14.000,29.895){2}{\rule{0.400pt}{0.507pt}}
\multiput(1160.58,281.00)(0.494,0.974){25}{\rule{0.119pt}{0.871pt}}
\multiput(1159.17,281.00)(14.000,25.191){2}{\rule{0.400pt}{0.436pt}}
\multiput(1174.58,305.33)(0.494,-0.680){25}{\rule{0.119pt}{0.643pt}}
\multiput(1173.17,306.67)(14.000,-17.666){2}{\rule{0.400pt}{0.321pt}}
\multiput(1188.00,287.93)(1.026,-0.485){11}{\rule{0.900pt}{0.117pt}}
\multiput(1188.00,288.17)(12.132,-7.000){2}{\rule{0.450pt}{0.400pt}}
\multiput(1202.58,282.00)(0.494,1.452){25}{\rule{0.119pt}{1.243pt}}
\multiput(1201.17,282.00)(14.000,37.420){2}{\rule{0.400pt}{0.621pt}}
\multiput(1216.58,322.00)(0.494,0.607){25}{\rule{0.119pt}{0.586pt}}
\multiput(1215.17,322.00)(14.000,15.784){2}{\rule{0.400pt}{0.293pt}}
\multiput(1230.58,331.94)(0.494,-2.039){25}{\rule{0.119pt}{1.700pt}}
\multiput(1229.17,335.47)(14.000,-52.472){2}{\rule{0.400pt}{0.850pt}}
\multiput(1244.58,275.47)(0.494,-2.186){25}{\rule{0.119pt}{1.814pt}}
\multiput(1243.17,279.23)(14.000,-56.234){2}{\rule{0.400pt}{0.907pt}}
\multiput(1258.00,223.59)(0.786,0.489){15}{\rule{0.722pt}{0.118pt}}
\multiput(1258.00,222.17)(12.501,9.000){2}{\rule{0.361pt}{0.400pt}}
\multiput(1272.58,232.00)(0.494,1.452){25}{\rule{0.119pt}{1.243pt}}
\multiput(1271.17,232.00)(14.000,37.420){2}{\rule{0.400pt}{0.621pt}}
\multiput(1286.00,272.58)(0.637,0.492){19}{\rule{0.609pt}{0.118pt}}
\multiput(1286.00,271.17)(12.736,11.000){2}{\rule{0.305pt}{0.400pt}}
\put(701.0,408.0){\rule[-0.200pt]{3.373pt}{0.400pt}}
\multiput(1314.58,283.00)(0.494,0.570){25}{\rule{0.119pt}{0.557pt}}
\multiput(1313.17,283.00)(14.000,14.844){2}{\rule{0.400pt}{0.279pt}}
\multiput(1328.00,299.59)(0.728,0.489){15}{\rule{0.678pt}{0.118pt}}
\multiput(1328.00,298.17)(11.593,9.000){2}{\rule{0.339pt}{0.400pt}}
\multiput(1341.58,305.33)(0.494,-0.680){25}{\rule{0.119pt}{0.643pt}}
\multiput(1340.17,306.67)(14.000,-17.666){2}{\rule{0.400pt}{0.321pt}}
\multiput(1355.58,284.55)(0.494,-1.231){25}{\rule{0.119pt}{1.071pt}}
\multiput(1354.17,286.78)(14.000,-31.776){2}{\rule{0.400pt}{0.536pt}}
\multiput(1369.58,250.67)(0.494,-1.195){25}{\rule{0.119pt}{1.043pt}}
\multiput(1368.17,252.84)(14.000,-30.835){2}{\rule{0.400pt}{0.521pt}}
\multiput(1383.58,219.69)(0.494,-0.570){25}{\rule{0.119pt}{0.557pt}}
\multiput(1382.17,220.84)(14.000,-14.844){2}{\rule{0.400pt}{0.279pt}}
\multiput(1397.58,206.00)(0.494,1.048){25}{\rule{0.119pt}{0.929pt}}
\multiput(1396.17,206.00)(14.000,27.073){2}{\rule{0.400pt}{0.464pt}}
\multiput(1411.58,235.00)(0.494,2.113){25}{\rule{0.119pt}{1.757pt}}
\multiput(1410.17,235.00)(14.000,54.353){2}{\rule{0.400pt}{0.879pt}}
\multiput(1425.58,293.00)(0.494,0.680){25}{\rule{0.119pt}{0.643pt}}
\multiput(1424.17,293.00)(14.000,17.666){2}{\rule{0.400pt}{0.321pt}}
\put(1300.0,283.0){\rule[-0.200pt]{3.373pt}{0.400pt}}
\end{picture}